\documentclass[a4paper]{article}
\usepackage{latexsym,amssymb,amsmath}
\usepackage{amscd,array,cite,dsfont,graphicx,mathtools,multirow,rotating,tensor,times,verbatim,youngtab}
\usepackage[bookmarksnumbered,bookmarksopen=true,bookmarksopenlevel=1,linktocpage,pdfstartview=FitH]{hyperref}
\usepackage[all]{hypcap}
\usepackage[margin=2cm]{geometry}

\newcolumntype{M}[1]{>{$}{#1}<{$}}

\DeclareMathAlphabet\Scr{U}{rsf}{m}{n} \makeatletter
\@addtoreset{equation}{section} \makeatother

\newcommand{\be}{\begin{equation}}
\newcommand{\ee}{\end{equation}}
\newcommand{\bea}{\begin{eqnarray}}
\newcommand{\eea}{\end{eqnarray}}
\newcommand{\ba}{\begin{array}}
\newcommand{\ea}{\end{array}}
\newcommand{\bit}{\begin{itemize}}
\newcommand{\eit}{\end{itemize}}
\newcommand{\ben}{\begin{enumerate}}
\newcommand{\een}{\end{enumerate}}

\allowdisplaybreaks[1]

\DeclareMathOperator{\tr}{tr}

\DeclareMathOperator{\Or}{O}
\DeclareMathOperator{\SO}{SO}
\DeclareMathOperator{\USp}{USp}
\DeclareMathOperator{\SL}{SL}
\DeclareMathOperator{\SU}{SU}
\DeclareMathOperator{\Un}{U}
\DeclareMathOperator{\Sp}{Sp}

\DeclareMathOperator{\SpO}{(S)O}

\newcommand{\R}{\mathds{R}}

\newcommand{\half}{\ensuremath{\tfrac{1}{2}}}
\newcommand{\STU}{\ensuremath{STU}}

\newcommand{\alg}[1]{\ensuremath{\mathfrak{#1}}}
\newcommand{\field}[1]{\ensuremath{\mathds{#1}}}
\newcommand{\rep}[1]{\ensuremath{\mathbf{#1}}}
\newcommand{\tyoung}{\tiny\young}

\newcommand{\ket}[1]{|#1\rangle}

\begin{document}

\begin{titlepage}
\begin{center}
\hfill Imperial/TP/2011/mjd/5\\
\hfill CERN-PH-TH/2011-004 \\

\vskip 1.5cm

{\huge \bf On the Black-Hole/Qubit Correspondence}

\vskip 1.5cm

{\bf L.~Borsten\,$^1$, M.~J.~Duff\,$^2$, A.~Marrani\,$^3$ and W.~Rubens\,$^2$}

\vskip 20pt

{\it ${}^1$ INFN Sezione di Torino \&  Dipartimento di Fisica Teorica,
Universit\`a di Torino \\
     Via Pietro Giuria 1, 10125 Torino, Italy}\\\vskip 5pt
     \texttt{borsten@to.infn.it}

    \vspace{10pt}

{\it $^2$ Theoretical Physics, Blackett Laboratory, Imperial College London,\\
 London SW7 2AZ, United Kingdom}\\\vskip 5pt
\texttt{m.duff@imperial.ac.uk}\\
\texttt{w.rubens@imperial.ac.uk}

     \vspace{10pt}

{\it ${}^3$ Physics Department, Theory Unit, CERN,\\
     CH -1211, Geneva 23, Switzerland}\\\vskip 5pt
     \texttt{Alessio.Marrani@cern.ch}

\end{center}

\vskip 2.2cm

\begin{center} {\bf ABSTRACT}\\[3ex]\end{center}
The entanglement classification of four qubits is related to the extremal black holes of the 4-dimensional $STU$ model via a time-like reduction to three dimensions. This correspondence is generalised to the entanglement classification of a very special four-way entanglement of eight qubits and the  black holes of the maximally supersymmetric $\mathcal{N}=8$ and exceptional magic $\mathcal{N}=2$ supergravity theories.




\vfill


\end{titlepage}

\newpage \setcounter{page}{1} \numberwithin{equation}{section}

\newpage\tableofcontents

\section{Introduction}

The advent of quantum theory heralded a new era of understanding - we inhabit a fundamentally probabilistic world founded upon the principle of quantum superposition. Since \emph{information} is stored, processed and distributed by physical phenomena, such a radical reassessment of reality ought to carry with it some profound implications for our theories of information and computation.   A concerted effort to understand these implications  swiftly developed into the fascinating and rapidly expanding  field of \emph{quantum information theory} (QIT) \cite{Nielsen:2000}. One of the principal goals of QIT is to characterise the behaviour and computational potential of information processing systems which utilise the fundamental properties of quantum mechanics.
There is an expectation that quantum theory may be exploited to perform computational tasks beyond the capability of any, even idealistic, purely classical device. This possibility enjoys a  certain poetry: just as the conventional  microchip  meets its fundamental limit, fixed by the onset  of quantum noise at the atomic scale,  the very same quantum phenomena open the door to new, superior,  modes of computation. A key component of today's quantum information toolkit is the quintessentially quantum phenomenon  of \emph{entanglement}. The quantum states of two or more entangled objects must be described with reference to each other, even though the individual objects may be spatially separated. This leads to classically inexplicable, but
experimentally observable, quantum correlations between the spatially separated systems - ``spooky'' action
at a distance as Einstein described it. Quantum entanglement is vital to the emerging technologies of quantum
computing, communication and cryptography. One of the longest standing open problems in QIT is a complete qualitative and quantitative
characterisation of multipartite entanglement.

In quite separate developments Black Holes (BHs) have commanded an equally privileged position in the various attempts to unify the fundamental interactions including quantum gravity. While general relativity refuses to succumb to quantum rule, BHs raise quandaries that strike at the very heart of quantum theory. Without a proper theory of quantum gravity, such paradoxes will continue to haunt us. M-theory, which grew out of pioneering work on supergravity and superstring theory, is a promising approach to quantum gravity. Living in eleven spacetime dimensions, it encompasses and connects the five consistent 10-dimensional superstring theories, as well as
11-dimensional supergravity and, as such, has the potential to unify the fundamental forces into a single consistent framework. However, M-theory is fundamentally non-perturbative and consequently remains largely mysterious, offering up only remote corners of its full structure.  The physics of BHs has occupied centre stage, providing unique insights into the non-perturbative structure of M-theory. Whatever final formulation M-theory eventually takes, understanding its BH solutions will play an essential role in its evolution.


For the most part these important endeavors in quantum information and gravity have led separate lives. However, the present work centres on a curious and unexpected interplay between these seemingly disparate themes. It constitutes one corner of the \emph{black-hole/qubit correspondence}: a relationship between the entanglement of qubits, the basic units of quantum information, and the entropy of BHs in M-theory. This story began in 2006 \cite{Duff:2006uz} when it was observed that the entropy of the $STU$ BH \cite{Duff:1995sm,Behrndt:1996hu,Bellucci:2008sv}, which appears in the compactification of M-theory to four dimensions, is given by \emph{Cayley's hyperdeterminant} \cite{Cayley:1845}. Remarkably, the   3-tangle \cite{Coffman:1999jd}, which measures the entanglement shared by three qubits, is also given by Cayley's hyperdeterminant \cite{Miyake:2002}. It was soon realised that there is in fact a one-to-one correspondence between the classification of 3-qubit entanglement \cite {Dur:2000} and the classification of  extremal \STU{} BHs \cite{Kallosh:2006zs}. Further work \cite{Levay:2006kf,Duff:2006ue,Levay:2006pt,Duff:2007wa,Levay:2007nm,Borsten:2008ur,Borsten:2008,Levay:2008mi,Borsten:2008wd,Levay:2009bp,Borsten:2009ae,Levay:2010qp,Levay:2010ua} has led to a more complete dictionary translating a variety of phenomena in one language to those in the other. It seems that we are, as yet, only glimpsing the tip of an iceberg.

Here we  develop further a recent application \cite{Borsten:2010db} of the black-hole/qubit correspondence  to the much more difficult problem of classifying 4-qubit entanglement. The experimental significance of this challenge  is re-enforced  as  4-qubit   entanglement   is now achievable in the laboratory \cite{Amselem:2009,Prevedel:2009,Lavoie:2010}.  The key technical ingredient is the Kostant-Sekiguchi theorem
\cite{Sekiguchi:1987,Collingwood:1993}, which provides the link between the BHs and qubits. Our main result, summarized in  \autoref{tab:realcosets}, is that there are 31 entanglement families which
reduce to nine up to permutations of the four qubits. Consulting  \autoref{tab:authors} we see that the nine agrees with \cite{Verstraete:2002,Chterental:2007}, while the 31 is new. From the BH perspective, we find that the attractor equations, which determine the amount of supersymmetry preserved by a particular BH solution, display a
symmetry consistent with permutations of the qubits. For example, the $A$-GHZ state yields a set of attractor equations which are invariant under a triality corresponding to the permutation of $B,C,D$ in the GHZ state.

We begin, in \autoref{sec:ent}, with an elementary introduction to
entanglement in QIT, with particular emphasis on \emph{Stochastic
Local Operations and Classical Communication} and the status of
4-qubit entanglement classification. In \autoref{sec:BH} we briefly
review BHs in supergravity and, in particular, the role of time-like
dimensional reduction and nilpotent orbits. In \autoref{sec:4qubit}
we invoke the Konstant-Sekiguchi theorem, which maps the BH
solutions to the 4-qubit entanglement classes and provide a detailed
analysis of both the structure of the BH solutions and the
entanglement classes. We conclude in Sects. \ref{sec:8qubit} and
\ref{sec:8qubit-exc} with the generalisation to $\mathcal{N}=8$ and
$\mathcal{N}=2$ exceptional supergravities, respectively; while
admitting a QIT interpretation as the four-way entanglement of eight
qubits, these theories are not amenable to the Kostant-Sekiguchi
theorem.

\section{Entanglement and SLOCC\label{sec:ent}}

In their seminal  1935 work Einstein, Podolsky, and Rosen (EPR) correctly concluded that assuming ``local realism'' the quantum mechanical wave function cannot provide a complete description of physical reality \cite{Einstein:1935rr}. Entanglement was identified as the chief culprit. They speculated on the existence of a more fundamental underlying (classical) theory that toed the line of local realism.   However, such questions remained a matter of philosophical preference, seemingly  inaccessible to experiment. All this changed in 1964 when  Bell introduced his now famous inequality \cite{Bell:1964kc}. In one fell swoop, entanglement had been elevated from a conceptual puzzle to an experimental observable confronting the very assumptions of local realism. This was Bell's great insight - to derive from EPR's  criteria something which could be used to check experimentally  the phenomenological viability of local realism.  Moreover, the Bell inequality  opened the door to utilising  entanglement in quantum information theoretic processes. For example,  it famously forms  the basis  of a secure cryptographic key distribution protocol  \cite{Ekert:1991}.

As quantum information theory developed, the role of entanglement became increasingly central. Entanglement  may be created, manipulated and consumed in the course of a given quantum computation or protocol. Futhermore,  it can in fact  exist in physically distinct forms. For example, multipartite states provide  so-called Bell inequalities without the inequality \cite{Greenberger:1990}. All this motivated a pressing need to properly quantify and classify entanglement. Conventionally, the state of a composite system is said to be entangled if it cannot be written as a tensor product of states of the constituent subsystems. However, this particular measure is perhaps insufficient to really capture the various subtleties of entanglement. For example, there are two totally non-separable 3-qubit states that have physically distinct entanglement properties \cite{Dur:2000}. Is there a more illuminating notion of entanglement? Let us take our cues from experiment. We do not actually observe the tensor product structure, even though it underpins our theoretical understanding. What we do observe are correlations between spatially separated systems that admit no classical explanation. This motivates the more general and quantum information theoretic notion of entanglement as  correlations between constituent pieces of a composite system that are of a quantum origin \cite{Bennett:1996,Bennett:1999,Plenio:2007}.
The question now is, how does one differentiate between classical correlations and those correlations which may be attributed to genuine quantum phenomena? Classical  correlations are \emph{defined} as those which  may be generated by \emph{Local Operations and Classical Communication} (LOCC) \cite{Bennett:1996,Bennett:1999,Plenio:2007}. Any classical correlation
may be experimentally established using LOCC. Conversely, all correlations unobtainable via LOCC are regarded as \emph{bona fide} quantum entanglement.

The LOCC paradigm is quite intuitive. Heuristically, given a
composite quantum system with its components spread among different
laboratories around the world, one allows each experimenter to
perform any quantum operation or measurement on their component
locally in their laboratory. These local operations cannot establish
any correlations, classical or quantum. However, the experimenters
may communicate any information they see fit via a classical channel
(carrier pigeon, smoke signals, e-mail). Any number of LO and CC
rounds may be performed. In this manner one may set-up arbitrary
classical correlations. However, since all information exchanged
between the separated parties at any point was intrinsically
classical, LOCC cannot create genuine quantum correlations.

Two quantum states of a composite system are then said to be \emph{%
stochastically} LOCC (SLOCC) equivalent if and only if they may be
probabilistically interrelated using LOCC. Since LOCC cannot create
entanglement, two SLOCC-equivalent states must possess the same
``amount'' of entanglement. For more details, see
\cite{Plenio:2007,Horodecki:2007} and Refs. therein.

Let us make this a little more precise by focusing on the specific
case of multi-qubit systems. What is a qubit? Quantum information
can live in a quantum mechanical superposition. Hence, the qubit is
a quantum superposition of the classical binary digits ``0'' and
``1''. The particular physical realisation (there are many: photon
polarisations, quantum dots, trapped ions, mode splitters, to name
but a few) of the qubit is not important, any two state quantum
system will do. Hence, qubits are simply
regarded abstractly as elements of the 2-dimensional Hilbert space $%
\mathds{C}^{2}$, equipped with the conventional norm, where the two
basis
states are labelled $|0\rangle $ and $|1\rangle $. An $n$-qubit bit string $%
|\Psi \rangle $ lives in the $n$-fold tensor product of
$\mathds{C}^{2}$:
\begin{equation}
\begin{split}
|\Psi \rangle & =a_{A_{1}\ldots A_{n}}|A_{1}\rangle \otimes
|A_{2}\rangle
\otimes \ldots |A_{n}\rangle \\
& =a_{A_{1}\ldots A_{n}}|A_{1}A_{2}\ldots A_{n}\rangle ,
\end{split}
\end{equation}
where $a_{A_{1}\ldots A_{n}}\in \mathds{C}$ and we sum over
$A_{1},\ldots
,A_{n}=0,1$. In \cite{Dur:2000} it was argued that two states of an $n$%
-qubit system are SLOCC-equivalent if and only if they are related by $[\SL%
(2,\mathds{C})]^{\otimes n}$, under which $a_{A_{1}\ldots A_{n}}$
transforms as the fundamental $\mathbf{(2,2,\ldots ,2)}$
representation. In this respect, $[\SL(2,\mathds{C})]^{\otimes n}$
may be usefully thought of as the ``gauge'' group of $n$-qubit
entanglement. Hence, the space of physically distinct $n$-qubit
entanglement classes (or orbits) is given by
\begin{equation}
\frac{{\mathds C}^{2}\otimes {\mathds C}^{2}\otimes \ldots \mathds{C}^{2}}{%
\SL_{1}(2,\mathds{C})\times \SL_{2}(2,\mathds{C})\times \ldots \SL_{n}(2,%
\mathds{C})}.
\end{equation}
When classifying entanglement, it is this space we wish to
understand.

This very quickly becomes a difficult task. Although two and three
qubit entanglement is well-understood (see \textit{e.g.}
\cite{Dur:2000}), the literature on four qubits can be confusing and
seemingly contradictory, as illustrated in \autoref{tab:authors}.
\begin{table*}[th]
\caption[Classification of $D=4,\ensuremath{\mathcal{N}}=2$
\ensuremath{STU}{} black holes.]{Various results on four-qubit
entanglement.} \label{tab:authors}
\begin{tabular*}{\textwidth}{@{\extracolsep{\fill}}clllr@{,}rr@{,}r}
\hline Paradigm & Author & Year & Ref & \multicolumn{2}{c}{result
mod perms} & \multicolumn{2}{c}{result incl. perms} \\ \hline
\multirow{5}{*}{classes} & Wallach & 2004 & \cite{Wallch:2008} &
\multicolumn{2}{c}{?} & \multicolumn{2}{c}{90} \\
& Lamata et al & 2006 & \cite{Lamata:2006b} & 8 genuine & 5
degenerate & 16
genuine & 18 degenerate \\
& Cao et al & 2007 & \cite{Cao:2007} & 8 genuine & 4 degenerate & 8
genuine
& 15 degenerate \\
& Li et al & 2007 & \cite{Li:2007c} & \multicolumn{2}{c}{?} &
$\geq31$
genuine & 18 degenerate \\
& Akhtarshenas et al & 2010 & \cite{Akhtarshenas:2010} &
\multicolumn{2}{c}{? } & 11 genuine & 6 degenerate \\
& Buniy et al & 2010 & \cite{Buniy:2010a} & 21 genuine & 5
degenerate & 64
genuine & 18 degenerate \\
\hline
\multirow{3}{*}{families} & Verstraete et al & 2002 &
\cite{Verstraete:2002}
& \multicolumn{2}{c}{9} & \multicolumn{2}{c}{?} \\
& Chterental et al & 2007 & \cite{Chterental:2007} &
\multicolumn{2}{c}{9} &
\multicolumn{2}{c}{?} \\
& String theory & 2010 & \cite{Borsten:2010db} &
\multicolumn{2}{c}{9} & \multicolumn{2}{c}{31} \\ \hline
\end{tabular*}
\end{table*}
This is due in part to genuine calculational disagreements, but in
part to the use of distinct (but in principle consistent and
complementary) perspectives on the criteria for classification.

On the one hand, there is the ``covariant'' approach which
distinguishes the SLOCC orbits by the vanishing or not of
$[\SL(2,\mathds{C})]^{\otimes n}$ covariants/invariants. This
philosophy is adopted for the 3-qubit case in
\cite{Dur:2000,Borsten:2009yb}, for example, where it was shown that
three qubits can be tripartite entangled in two inequivalent ways,
denoted $W$ and GHZ (Greenberger-Horne-Zeilinger). The analogous
4-qubit case was treated, with partial results, in
\cite{Briand:2003a}. Several new systems, in addition to the 4-qubit example, have been studied using the covariant framework in interesting recent work employing algebraic invariants of linear maps \cite{Buniy:2010b,Buniy:2010a}.

On the other hand, there is the ``normal form'' approach which
considers ``families'' of orbits. An arbitrary state may be
transformed into one of a finite number of normal forms. If the
normal form depends on some of the algebraically independent SLOCC
invariants it constitutes a family of orbits parametrised by these
invariants. On the other hand, a parameter-independent family
contains a single orbit. This philosophy is adopted for the 4-qubit
case $|\Psi \rangle =a_{ABCD}|ABCD\rangle $ in \cite
{Verstraete:2002,Chterental:2007}. Up to permutation of the four
qubits,
these authors found 6 parameter-dependent families called $G_{abcd}$, $%
L_{abc_{2}}$, $L_{a_{2}b_{2}}$, $L_{a_{2}0_{3\oplus \bar{1}}}$, $L_{ab_{3}}$%
, $L_{a_{4}}$ and 3 parameter-independent families called
$L_{0_{3\oplus \bar{1}}0_{3\oplus \bar{1}}}$, $L_{0_{5\oplus
\bar{3}}}$, $L_{0_{7\oplus \bar{1}}}$. For example, a family of
orbits parametrised by all four of the
algebraically independent SLOCC invariants is given by the normal form ${%
G_{abcd}}$:
\begin{equation}
\begin{split}
& \frac{\left( a+d\right) }{2}(|0000\rangle +|1111\rangle
)+\frac{\left(
a-d\right) }{2}(|0011\rangle +|1100\rangle ) \\
& +\frac{\left( b+c\right) }{2}(|0101\rangle +|1010\rangle
)+\frac{\left( b-c\right) }{2}(|1001\rangle +|0110\rangle ).
\end{split}
\label{eq:gabcd}
\end{equation}

To illustrate the difference between these two approaches, consider
the separable EPR-EPR state $(|00\rangle +|11\rangle )\otimes
(|00\rangle
+|11\rangle )$. Since this is obtained by setting $b=c=d=0$ in %
\eqref{eq:gabcd}, it belongs to the $G_{abcd}$ family, whereas in
the
covariant approach it forms its own class. Similarly, a totally separable $A$%
-$B$-$C$-$D$ state, such as $|0000\rangle $, for which all
covariants/invariants vanish, belongs to the family $L_{abc_{2}}$,
which however also contains genuine four-way entangled states. These
interpretational differences were also noted in \cite{Lamata:2006b}.

As we shall see, our BH perspective lends itself naturally to the
``normal form'' framework.

\section{Black Holes and Nilpotent Orbits \label{sec:BH}}

\subsection{Time-like Reduction and Stationary Black Holes}

We consider $D=4$ supergravity theories coupled to $n$ Abelian gauge
potentials, in which the scalar fields coordinatise  a
symmetric coset $M_{4}={G_{4}}/{H_{4}}$, where $G_{4}$ is the global
U-duality group\footnote{For a recent review of the general theory of duality rotations in
four-dimensional supergravity theories see \emph{e.g.} \cite{Aschieri:2008ns}.} and $H_{4}$ is its maximal compact subgroup. In this
paper we will consider both the $\ensuremath{\mathcal{N}}=2$ $STU$
supergravity coupled to three vector multiplets, for which $n=4$,
and the full $\ensuremath{\mathcal{N}}=8$ theory, for which $n=28$. Schematically, we
have a action of the form
\begin{equation}
S=\int d^{4}x\sqrt{-g}\left[ R_{4}-\partial _{\mu }\phi _{i}\partial
^{\mu }\phi _{j}\gamma _{4}^{ij}-\mu _{IJ}F^{I}\wedge ^{{}}\star
F^{J}+\nu _{IJ}F^{I}\wedge F^{J}\right]
\end{equation}
where $R_{4}$ is the Ricci scalar, $\phi ^{i}$ are the scalar fields
(coordinates in $M_{4}$), $\gamma _{4}^{ij}$ is the $M_{4}$ metric
and $F^{I} $ are the $n$ field strengths of the $n$ Abelian gauge
vectors. We are going to use a spherically symmetric static
\textit{Ansatz} of the form
\begin{equation}
ds^{2}=-e^{U}dt^{2}+e^{-U}(dr^{2}+r^{2}d\Omega
^{2})\label{BH-metric-D=4}
\end{equation}
to describe our BH background. If we were to compactify one of
the space like directions, we would end up with a $D=3$ theory of
spacetime with a scalar manifold given by $M_{3}=G_{3}/H_{3}$ where
$G_{3}$ is the $D=3$ duality group and $H_{3}$ is its maximal
compact subgroup. Instead (for reasons that will become clear), we
perform a time-like reduction to a $D=3$ space (not spacetime)
$\Sigma _{3}$. In $D=3$ all vectors can be dualised to scalars, such
that after dualisation one ends up with a non-linear sigma model
coupled to Euclidean gravity, \textit{i.e.}\thinspace\ an action of
the form \cite{Breitenlohner:1987dg}
\begin{equation}
S=\int d^{3}x\sqrt{h}\left[
\frac{1}{2}R_{3}-\frac{1}{2}h^{ab}\partial _{a}\phi ^{i}\partial
_{b}\phi ^{j}\gamma _{3ij}\right]
\end{equation}
where $h_{ab}$ is the (Euclidean) $D=3$ metric of $\Sigma _{3}$,
$R_{3}$ is the Ricci scalar, $\phi ^{i}$ are $D=3$ scalars
coordinatizing $M_{3}^{\ast }
$ (which is now a pseudo-Riemannian symmetric space $M_{3}^{\ast }={G_{3}}/{%
H_{3}^{\ast }}$, where $H_{3}^{\ast }$ is a suitable non-compact form of $%
H_{3}$), and $\gamma _{3ij}$ is the $M_{3}^{\ast }$ metric. One may
wonder if this procedure is well defined and what the pay-off might
be. Luckily, as explained in \cite{Breitenlohner:1987dg}, as long as
one considers stationary BHs with a well defined global time-like
Killing vector (in the original $D=4$) the mentioned procedure is
well defined.

The equations of motion are
\begin{align}
R_{ab}& =\gamma _{3ij}\partial _{a}\phi ^{i}\partial _{b}\phi ^{j}; \\
D^{\alpha }\partial _{\alpha }\phi ^{i}& =0.
\end{align}
With a judicious coordinate change to $\tau =r^{-1}$, the equations
of
motion for the scalars and gravity decouple, and reduce to geodesics on $%
M_{3}^{\ast }$ that are parametrised by $\tau $, given by
\begin{equation}
\frac{d^{2}\phi ^{i}}{d\tau ^{2}}+\Gamma _{jk}^{i}\frac{d\phi ^{j}}{d\tau }%
\frac{d\phi ^{k}}{d\tau }=0.
\end{equation}
Physically, integrating out one of these geodesics corresponds to
integrating the BH solution from $r=\infty $ to $r=0$ at the
horizon. These geodesics can be calculated from a Lagrangian of the
form
\begin{equation}
\mathcal{L}=\frac{1}{2}\gamma _{3ij}\dot{\phi}^{i}\dot{\phi}^{j}\label%
{Lagr-1}
\end{equation}
where the dots denote differentiation with respect to $\tau $. The
differential geometry of symmetric manifolds can thus be exploited
in order to re-express Lagrangian (\ref{Lagr-1}) in terms of Lie
algebra elements. Actully, the procedure under consideration moved
the time coordinate (and the $g_{tt}$ component of the metric) into
the pseudo-Riemannian scalar manifold $M_{3}^{\ast }$ itself.

The pay-off from this procedure is that the differential geometry
tools associated with symmetric manifolds can be used to study
properties of the BH solutions associated with the $g_{tt}$
component of the original $D=4$ metric (\ref{BH-metric-D=4}). In
particular, simple requirements, such as regularity and the type of
geodesic curves, allow one to select stationary and extremal BHs.
The Hamiltonian constraint is
\begin{equation}
\gamma _{3ij}\dot{\phi}^{i}\dot{\phi}^{j}=v^{2}.
\end{equation}
Given that $M_{3}^{\ast }$ is pseudo-Riemannian, the geodesics may
be time-like, light-like or space-like according as the solution is
non-extremal, extremal or over-extremal (unphysical).

One uses a coset representative $L\in G/H^{\ast }$, which transforms
globally under $G$ and locally under $H^{\ast }$ as
\begin{align}
L& \rightarrow gL\quad g\in G; \\
L& \rightarrow Lh\quad h\in H^{\ast }.
\end{align}
The vielbein and connection one-forms may be found in the
Maurer-Cartan formula $L^{-1}dL\in \ensuremath{\mathfrak{g}}$
\begin{equation}
L^{{-1}}dL=d\phi ^{i}V_{i}^{A}T_{A}
\end{equation}
where the $T_{A}\in \ensuremath{\mathfrak{g}}$ are in the solvable (\textit{%
i.e.}\ upper-triangular) parametrisation. One then further defines
the symmetric matrix $M=L\eta L^{T}$ such that $M$ transforms under
global $G$ in the adjoint as $M\rightarrow gMg^{-1}$ and is inert
under local $H^{\ast }
$ (intuitively, $M$ can be thought as a point on the manifold $G/H^{\ast }$%
). Thus, the following result is achieved
\cite{Breitenlohner:1987dg}:
\begin{align}
S& =\int \frac{1}{2}\gamma _{ij}\dot{\phi}^{i}\dot{\phi}^{j}=\int \frac{1}{2}%
\gamma _{ij}V_{A}^{i}V_{B}^{j}\dot{\phi}^{A}\dot{\phi}^{B}=\int \frac{1}{2}%
\eta _{AB}\dot{\phi}^{A}\dot{\phi}^{B} \\
& =\int \frac{1}{2}\tr(T_{A}T_{B})\dot{\phi}^{A}\dot{\phi}^{B}=\int \frac{1}{%
8}\tr(\dot{M}\dot{M}^{-1}),
\end{align}
where the last line takes about half a page of calculation to
manipulate the $\phi _{A}T^{A}$ into $M$. From here, the equation of
motion is clearly seen as
\begin{equation}
\frac{d}{d\tau }\left[ M^{-1}\frac{d}{d\tau }M\right] ,
\end{equation}
and the solution is given simply by
\begin{equation}
M(\tau )\equiv M(\phi ^{i}(\tau ))=M(0)\exp {2Q\tau }.\label{M-sol}
\end{equation}
where $Q\in \ensuremath{\mathfrak{g}}$ is an algebra element that
will become central to the whole picture explained below.

Theorem 6.4 of \cite{Breitenlohner:1998cv} states that \emph{any}
static spherically symmetric BHs is $G_{3}$-equivalent to the
Schwarzschild one in which the only scalar turned on is $g_{tt}=\exp
{U}$. Thus, one can study BH solutions by analyzing the orbits of
$M$ under $G_{3}$. In turns out that
\cite{Bergshoeff:2008be,Bossard:2009we,Bossard:2009at,Levay:2010ua},
in the adjoint, the Lie algebra valued matrix of $D=3$ Noether
charges $Q$ satisfies
\begin{equation}
Q^{5}=5v^{2}Q^{3}-4v^{2}Q^{2},
\end{equation}
while in the fundamental it holds that
\begin{equation}
Q^{3}=v^{2}Q
\end{equation}
where $v^{2}$ is the geodesic parameter. Therefore, for light-like
geodesics (where $v^{2}=0$), corresponding to extremal BHs, one
obtains that $Q^{3}=0$ is nilpotent. From (\ref{M-sol}), this
implies that $M$ terminates at
\begin{equation}
M=\left[ \ensuremath{\mathds{I}}+\tau Q+\frac{1}{2}\tau
^{2}Q^{2}\right] ,
\end{equation}
and that the problem of classifying extremal BH solutions reduces to
the problem of classifying orbits of nilpotent $Q\in
\ensuremath{\mathfrak{g}}$ or, in other words, the nilpotent orbits
of $G_{3}$ given are in one-to-one correspondence with the extremal
BHs of the original $D=4$ theory.

By specializing the above reasoning to the $STU$ model, one
gets\footnote{The \textit{rank} of a globally symmetric space is
defined as the maximal dimension (in $\mathds{R}$) of a
\textit{flat} (\textit{i.e.} with vanishing Riemann tensor),
\textit{totally geodesic} submanifold of such a space (see
\textit{e.g.} $\S 6$, page 209 of \cite{Helgason:1978}).}
\begin{gather}
\frac{G_{4}}{H_{4}}=\frac{\SL(2,\ensuremath{\mathds{R}})\times \SL(2,%
\ensuremath{\mathds{R}})\times
\SL(2,\ensuremath{\mathds{R}})}{\SO(2)\times
\SO(2)\times \SO(2)}, \text{rank}=3;\label{STU-D=4-1} \\
\frac{G_{3}}{H_{3}^{\ast }}=\frac{\SO(4,4)}{\SO(2,2)\times \SO(2,2)}, \text{rank}=4.\label%
{STU-D=3-1}
\end{gather}
Whereas for the maximal $\ensuremath{\mathcal{N}}=8$ theory it holds
\begin{gather}
\frac{G_{4}}{H_{4}}=\frac{E_{7(7)}}{\SU(8)}, \text{rank}=7;  \label{N=8-D=4-1} \\
\frac{G_{3}}{H_{3}^{\ast }}=\frac{E_{8(8)}}{\SO^{\ast }(16)},
\text{rank}=8. \label{N=8-D=3-1}
\end{gather}
In $\mathcal{N}=2$ theories, the relation between the special
K\"{a}hler (see \textit{e.g.}
\cite{Ceresole:1995ca,Andrianopoli:1996cm} and Refs. therein)
symmetric coset (\ref{STU-D=4-1}) and the para-quaternionic symmetric coset (%
\ref{STU-D=4-1}) is mathematically expressed through the ``$\ast
$-version'' of the $c$-map \cite{Cecotti:1988qn} (see also
\cite{Bergshoeff:2008be}, and Refs. therein).

In the case of the $STU$ model, thanks to the Kostant-Sekiguchi
correspondence (see next Subsection), the nilpotent orbits of
$G_{3}/H_{3}^{\ast }$ are diffeomorphic to the complex nilpotent
orbits of $[\SL(2,\ensuremath{\mathds{C}})]^{4}$ on its fundamental,
which happens to be the classification of four qubits, see the
treatment in Sec. \ref{sec:4qubit}.




\subsection{The Kostant-Sekiguchi Theorem}

Consider a complex Lie algebra
$\ensuremath{\mathfrak{g}}_{\field{C}}$ with
Cartan decomposition $\ensuremath{\mathfrak{g}}_{\field{C}}=%
\ensuremath{\mathfrak{h}}_{\ensuremath{\mathds{C}}}+\ensuremath{\mathfrak{k}}%
_{\ensuremath{\mathds{C}}}$. It holds that
\begin{equation}
\lbrack \ensuremath{\mathfrak{h}}_{\mathds{C}},\ensuremath{\mathfrak{k}}_{%
\mathds{C}}]\subseteq \ensuremath{\mathfrak{k}}_{\mathds{C}},\quad \lbrack %
\ensuremath{\mathfrak{k}}_{\mathds{C}},\ensuremath{\mathfrak{k}}_{\mathds{C}%
}]\subseteq \ensuremath{\mathfrak{h}}_{\mathds{C}},\quad \lbrack %
\ensuremath{\mathfrak{h}}_{\mathds{C}},\ensuremath{\mathfrak{h}}_{\mathds{C}%
}]\subseteq \ensuremath{\mathfrak{h}}_{\mathds{C}},
\end{equation}
(\textit{i.e.}\ $\ensuremath{\mathfrak{h}}_{\mathds{C}}$ is a
sub-algebra of
$\ensuremath{\mathfrak{g}}_{\mathds{C}}$). The $\ensuremath{\mathfrak{g}}_{%
\mathds{C}}$ and $\ensuremath{\mathfrak{h}}_{\mathds{C}}$ algebras,
have corresponding complex Lie groups, $G_{\mathds{C}}$ and
$H_{\mathds{C}}$, that have a natural adjoint action on their
respective algebras, given by
\begin{equation}
\ensuremath{\mathfrak{a}}\rightarrow
g\ensuremath{\mathfrak{a}}g^{-1}
\end{equation}
where $\ensuremath{\mathfrak{a}}\in \ensuremath{\mathfrak{g}}$ and $g\in G_{%
\mathds{C}}$. Consider further the real forms of $\ensuremath{\mathfrak{g}}_{%
\mathds{C}}$ and $\ensuremath{\mathfrak{h}}_{\mathds{C}}$ given
respectively
by $\ensuremath{\mathfrak{g}}_{\mathds{R}}$ and $\ensuremath{\mathfrak{h}}_{%
\mathds{R}}$ and their respective real groups $G_{\mathds{R}}$ and $H_{%
\mathds{R}}$. Then the Kostant-Sekiguchi theorem
\cite{Collingwood:1993}
states that the adjoint orbits of $G_{\mathds{R}}$ on elements of $%
\ensuremath{\mathfrak{g}}_{\mathds{R}}$ that are nilpotent are are
diffeomorphic to the nilpotent fundamental orbits of $H_{\mathds{C}}$ on $%
\ensuremath{\mathfrak{k}}_{\mathds{C}}$, \textit{i.e.}\
\begin{equation}
\frac{\ensuremath{\mathfrak{N}}\cap \ensuremath{\mathfrak{g}}_{\mathds{R}}}{%
G_{\mathds{R}}}\leftrightarrow \frac{\ensuremath{\mathfrak{N}}\cap %
\ensuremath{\mathfrak{k}}_{\mathds{C}}}{H_{\mathds{C}}}.
\end{equation}
where $\ensuremath{\mathfrak{N}}$ is the variety of nilpotent
elements.

For the $STU$ model, we pick
\begin{align}
\ensuremath{\mathfrak{g}}_{\field{C}}=\ensuremath{\mathfrak{so}}(8)_{%
\field{C}}& =\ensuremath{\mathfrak{so}}(4)_{\field{C}}+\ensuremath{%
\mathfrak{so}}(4)_{\field{C}}+\ensuremath{\mathbf{(4,4)}} \\
& =\ensuremath{\mathfrak{sl}}(2)_{\field{C}}+\ensuremath{\mathfrak{sl}}(2)_{%
\field{C}}+\ensuremath{\mathfrak{sl}}(2)_{\field{C}}+\ensuremath{%
\mathfrak{sl}}(2)_{\field{C}}+\ensuremath{\mathbf{(2,2,2,2)}} \\
& =\ensuremath{\mathfrak{h}}_{\ensuremath{\mathds{C}}}+\ensuremath{%
\mathfrak{k}}_{\ensuremath{\mathds{C}}},
\end{align}
therefore, we have $H_{\mathds{C}}=\SL(2,\mathds{C})\times \SL(2,\mathds{C})\times \SL(2,%
\mathds{C})\times \SL(2,\mathds{C})$ and $\ensuremath{\mathfrak{k}}_{C}=%
\ensuremath{\mathbf{(2,2,2,2)}}_{\mathds{C}}$. We choose
$G_{\mathds{R}}=\SO_0(4,4)$, where the $0$ subscript denotes the
identity-connected component, and pick the non-compact version of the maximal
compact subgroup $H_{\mathds{R}}^{\ast
}=\SO(2,2)\times \SO(2,2)$\footnote{%
The Kostant-Sekiguchi theorem applies to non-compact $H_{\mathds{R}}^{*}$ as
well. The details are in the first appendix of
\cite{Bossard:2009we}}. In this way the Kostant-Sekiguchi
correspondence tells us that
\begin{equation}
\text{Nilpotent}\frac{\ensuremath{\mathfrak{so}}(4,4)}{\SO(4,4)}\sim \text{%
Nilpotent}\frac{\ensuremath{\mathbf{(2,2,2,2)}}_{\mathds{C}}}{\SL(2,\mathds{C}%
)^{4}}.
\end{equation}
while for the $\ensuremath{\mathcal{N}}=8$ supergravity and
$\ensuremath{\mathcal{N}}=2$ exceptional model, we choose
\begin{align}
\ensuremath{\mathfrak{g}}_{\field{C}}=\ensuremath{\mathfrak{e}}(8)_{\field{C}%
}& =\ensuremath{\mathfrak{so}}^{*}(16)_{\field{C}}+\ensuremath{\mathbf{(128)}} \\
& =\ensuremath{\mathfrak{h}}_{\ensuremath{\mathds{C}}}+\ensuremath{%
\mathfrak{k}}_{\ensuremath{\mathds{C}}}.
\end{align}
However, as it will be discussed in Sects. \ref{sec:8qubit} and \ref{sec:8qubit-exc}, the QIT interpretations of both the $%
\ensuremath{\mathcal{N}}=8$ and the exceptional $\ensuremath{\mathcal{N}}=2$
 theories are not amenable to the application of the
Kostant-Sekiguchi correspondence.



\section{The $STU$ model and the Entanglement of Four Qubits \label%
{sec:4qubit}}

\subsection{Summary}

Here we briefly summarise the relationship between the classes of
$STU$ BH solutions and the entanglement classes of four qubits. In
the following Section we provide a more comprehensive analysis.

The $STU$ model \cite
{Duff:1995sm,Behrndt:1996hu,Gimon:2007mh,Bellucci:2008sv} is a
particular model of $\ensuremath{\mathcal{N}}=2$ supergravity
coupled to three vector multiplets. It has three complex scalars
denoted $S,T$ and $U$, which parameterize the symmetric coset space
(\ref{STU-D=4-1}).

The static, asymptotically flat, spherically symmetric, extremal\footnote{%
BHs are divided into extremal and non-extremal, according as their
Hawking temperature is zero or not. The corresponding Noether orbits
in $D=3$ are nilpotent or semisimple, respectively. } BH solutions
of the $STU$ model are characterized by a maximum of 8 charges (four
electric and four magnetic), namely $1+3$ from the gravity and
vector multiplets respectively, plus their magnetic duals. Hence,
the Bekenstein-Hawking entropy
\cite{Bekenstein:1973ur,Bardeen:1973gs} is a function of the 8
charges. Through scalar-dressing, these
charges can be grouped into the $\ensuremath{\mathcal{N}}=2$ central charge $%
z$ and three ``matter charges''. Depending on the values of the
charges, the extremal BHs are divided into ``small'' or ``large'',
according as their Bekenstein-Hawking
\cite{Bekenstein:1973ur,Bardeen:1973gs} entropy is zero or not. The
``small'' ones are termed \textit{lightlike}, \textit{critical}
or \textit{doubly-critical}, depending on the minimal number (under $U$%
-duality) of representative electric or magnetic charges, which is
respectively 3, 2 or 1. One subtlety is that some extremal cases,
termed ``extremal'', cannot be obtained as limits of non-extremal
BHs (see Sect. \ref{"extr"}).

Performing a time-like reduction to $D=3$, one obtains the
pseudo-Riemannian para-quaternionic symmetric coset
(\ref{STU-D=3-1}). Hence, the extremal solutions are classified by
the nilpotent orbits of $\SO(4,4)$ acting on its adjoint
representation $\mathbf{28}$. Here we consider the finer
classification, obtained from the nilpotent orbits of
$\SO_{0}(4,4)$. These orbits may be labeled by
``signed'' Young tableaux, often referred to as $ab$-diagrams in the
mathematics literature (see \textit{e.g.} \cite{Djokovic:2000}, and
Refs. therein). Each signed Young tableau, as listed in %
\autoref{tab:realcosets}, actually corresponds to a single nilpotent $\Or%
(4,4)$ orbit, of which the $\SO_{0}(4,4)$ nilpotent orbits are the
connected
components. Since $\Or(4,4)$ has four components, for each nilpotent $\Or%
(4,4)$ orbit there may be either 1, 2 or 4 nilpotent $\SO_{0}(4,4)$
orbits. This number is also determined by the corresponding signed
Young tableau. If the middle sign of every odd length row is ``$-$''
(``$+$'') there are 2 orbits and we label the diagram to its left
(right) with a $I$ or a $II$. If it only has even length rows, there
are 4 orbits and we label the diagram to both its left and right
with a $I$ or a $II$. If it is none of these, it is said to be
stable and there is only one orbit. The signed Young tableaux
together with their labellings, as listed in
\autoref{tab:realcosets}, give a total of 31 nilpotent
$\SO_{0}(4,4)$ orbits \cite{Borsten:2010db}. The
matching of the extremal classes to the nilpotent orbits is given in %
\autoref{tab:realcosets} \cite{Borsten:2010db}, and it is discussed
in detail in Sects. \ref{Large-Extremal}-\ref{"extr"}. We also
supply the complete list of the associated cosets in
\autoref{tab:realcosets}, some of which may be found in
\cite{Bossard:2009we}.

In order to relate the extremal BH solutions to the entanglement
classes of four qubits, we invoke the aforementioned
Kostant-Sekiguchi theorem \cite
{Sekiguchi:1987,Collingwood:1993}. Noting the convenient isomorphism $\SO%
(2,2)\cong \SL(2,\mathds{R})\times \SL(2,\mathds{R})$, the scalar manifold $%
G_{3}/H_{3}^{\ast }$ of the time-like reduced $STU$ model may be
rewritten as $\SO(4,4)/[\SL(2,\ensuremath{\mathds{R}})]^{4}$, which
yields the Cartan decomposition
\begin{equation}
\ensuremath{\mathfrak{so}}(4,4)\cong \lbrack \ensuremath{\mathfrak{sl}}(2,%
\ensuremath{\mathds{R}})]^{4}\oplus \ensuremath{\mathbf{(2,2,2,2)}}.
\label{eq:4qubitKS}
\end{equation}
The relevance of \eqref{eq:4qubitKS} to four qubits was pointed out
in \cite {Borsten:2008wd} and recently spelled out more clearly by
Levay \cite {Levay:2010ua}, who relates four qubits to $D=4$ $STU$
BHs. The 16 independent components are given by the $4+4$
electromagnetic charges, the NUT charge, the mass and three complex
scalars of the $STU$ model. By
applying the Kostant-Sekiguchi correspondence to the Cartan decomposition (%
\ref{eq:4qubitKS}), one can state that the nilpotent orbits of
$\SO_{0}(4,4)$ acting on its adjoint representation are in
one-to-one correspondence with the nilpotent orbits of
$[\SL(2,\ensuremath{\mathds{C}})]^{4}$ acting on its fundamental
$\ensuremath{\mathbf{(2,2,2,2)}}$ representation and, hence, with
the classification of 4-qubit entanglement. Note furthermore that it
is the complex qubits that appear automatically, thereby relaxing
the restriction to real qubits (sometimes called rebits) that
featured in earlier versions of the BH/qubit correspondence.

It follows that there are 31 nilpotent orbits for four qubits under
SLOCC \cite{Borsten:2010db}. For each nilpotent orbit there is
precisely one family of SLOCC orbits since each family contains one
nilpotent orbit on setting all invariants to zero. The nilpotent
orbits and their associated families are summarized in
\autoref{tab:realcosets} \cite{Borsten:2010db}, which is split into
upper and lower sections according as the nilpotent orbits belong to
parameter-dependent or parameter-independent families.

If one allows for the permutation of the four qubits, the connected
components of each $\Or(4,4)$ orbit are re-identified reducing the
count to 17. Moreover, these 17 are further grouped under this
permutation symmetry into just nine nilpotent orbits. It is not
difficult to show that these nine cosets match the nine families of
\cite{Verstraete:2002,Chterental:2007}, as listed in the final
column of \autoref{tab:realcosets} (provided we adopt the version of
$L_{ab_{3}}$ presented in \cite{Chterental:2007} rather than
in \cite{Verstraete:2002}). For example, the state representative $%
|0111\rangle +|0000\rangle $ of the family $L_{0_{3\oplus
\bar{1}}0_{3\oplus
\bar{1}}}$ is left invariant by the $[\SO(2,\ensuremath{\mathds{C}}%
)]^{2}\times \ensuremath{\mathds{C}}$ subgroup, where $[\SO(2,%
\ensuremath{\mathds{C}})]^{2}$ is the stabilizer of the three-qubit
GHZ
state \cite{Borsten:2009yb}. In contrast, the four-way entangled family $%
L_{0_{7\oplus \bar{1}}}$, which is the ``principal'' nilpotent orbit
\cite {Collingwood:1993}, is not left invariant by any subgroup.
Note that the total of 31 does not follow trivially by permuting the
qubits in these nine. Naive permutation produces far more than 31
candidates, which then have to be reduced to SLOCC inequivalent
families.

There is a satisfying consistency of this process with respect to
the covariant approach (which, as mentioned, is the other criterion
for classification). For example, the covariant classification has
four biseparable classes $A$-GHZ, $B$-GHZ, $C$-GHZ and $D$-GHZ which
are then identified as a single class under the permutation
symmetry. These four
classes are in fact the four nilpotent orbits corresponding to the families $%
L_{0_{3\oplus \bar{1}}0_{3\oplus \bar{1}}}$ in
\autoref{tab:realcosets}, which are also identified as a single
nilpotent orbit under permutations. Similarly, each of the four
$A$-W classes is a nilpotent orbit belonging to one of the four
families labeled $L_{a_{2}0_{3\oplus \bar{1}}}$ which are again
identified under permutations. A less trivial example is given by
the six $A$-$B$-EPR classes of the covariant classification. These
all lie in the single family $L_{a_{2}b_{2}}$ of
\cite{Verstraete:2002}, which is defined up to permutation.
Consulting \autoref{tab:realcosets} we see that, when not allowing
permutations, this family splits into six pieces, each containing
one of the six $A$-$B$-EPR classes. Finally, the single totally
separable class $A$-$B$-$C$-$D$ is the single nilpotent orbit inside
the single family $L_{abc_{2}}$, which maps into itself under
permutations.
\begingroup
\newlength\BHCol
\setlength\BHCol{1.4cm}
\newlength\RepCol
\setlength\RepCol{2.4cm}
\setlength{\tabcolsep}{1pt}

\begin{sidewaystable}
\small
\caption{Each black hole nilpotent $\SO_0(4,4)$ orbit  corresponds
to a 4-qubit nilpotent $[\SL(2,\field{C})]^4$ orbit
\cite{Borsten:2010db} \label{tab:realcosets}}
\begin{tabular*}{\textwidth}{@{\extracolsep{\fill}}>{\centering}m{\BHCol}*{4}{M{c}}>{\centering$}m{\RepCol}<{$}cM{c}}
\hline\hline
\multicolumn{3}{c}{$STU$ black holes}                                                                                                                                                                                                                                                                          & \multirow{2}{*}{$\dim_\field{R}$} & \multicolumn{4}{c}{Four qubits}                                                                                                                                                                                                   \\
\cline{1-3}\cline{5-8}\\[-6pt]
description                           & \text{Young tableaux}                                                                                                                       & \SO_0(4,4)\text{ coset}                                                                                                  &                                   & [\SL(2,\field{C})]^4\text{ coset}                                                                        & \text{nilpotent rep}                                            &        & \text{family}                               \\
\\[-6pt]\hline\hline\\[-6pt]
trivial                               & \text{trivial}                                                                                                                              & \frac{\SO_0(4,4)}{\SO_0(4,4)}                                                                                            & 1                                 & \frac{[\SL(2,\field{C})]^4}{[\SL(2,\field{C})]^4}                                                        & 0                                                               & $\in$  & G_{abcd}                                    \\
\\[-6pt]\hline\\[-6pt]
doubly-critical \half BPS             & \tyoung(+-,-+,+,-,+,-)                                                                                                                      & \frac{\SO_0(4,4)}{[\SL(2,\field{R})\times \SO(2,2,\field{R})]\ltimes[(\rep{2,4})^{(1)} \oplus{\rep1}^{(2)}]}             & 10                                & \frac{[\SL(2,\field{C})]^4}{[\SO(2,\field{C})]^3\ltimes\field{C}^4}                                      & \ket{0110}                                                      & $\in$  & L_{abc_2}                                   \\
\\[-6pt]\hline\\[-6pt]
                                      & \tyoung(-+-,+,-,+,-,+)                                                                                                                      & \frac{\SO_0(4,4)}{\SO(3,2;\field{R})\ltimes[(\rep{5\oplus1})^{(2)}]}                                                     &                                   &                                                                                                          &                                                                 &        &                                             \\
\\[-6pt]
critical, \half BPS and non-BPS       & \tyoung(+-+,-,+,-,+,-)                                                                                                                      & \frac{\SO_0(4,4)}{\SO(2,3;\field{R})\ltimes[(\rep{5\oplus1})^{(2)}]}                                                     & 12                                & \frac{[\SL(2,\field{C})]^4}{[\SO(3,\field{C})\times \field{C}]\times[\SO(2,\field{C})\ltimes \field{C}]} & \ket{0110}+\ket{0011}                                           & $\in$  & L_{a_2b_2}                                  \\
\\[-6pt]
                                      & \begin{pmatrix}I,II&\tyoung(+-,-+,+-,-+)&I,II\end{pmatrix}                                                                                  & \frac{\SO_0(4,4)}{\Sp(4,\field{R})\ltimes[(\rep{5\oplus1})^{(2)}]}                                                       &                                   &                                                                                                          &                                                                 &        &                                             \\
\\[-6pt]\hline\\[-6pt]
lightlike \half BPS and non-BPS       & \begin{array}{c}\begin{pmatrix}I,II&\tyoung(+-+,-+,+-,-)\;\end{pmatrix}\\\begin{pmatrix}\;\tyoung(-+-,+-,-+,+)&I,II\end{pmatrix}\end{array} & \frac{\SO_0(4,4)}{\SL(2,\field{R})\ltimes[(2\times {\rep 2})^{(1)}\oplus(3\times{\rep 1})^{(2)}\oplus{\rep 2}^{(3)}]}    & 16                                & \frac{[\SL(2,\field{C})]^4}{[\SO(2,\field{C})\ltimes\field{C}]\times \field{C}^2}                        & \ket{0110}+\ket{0101}+\ket{0011}                                & $\in$  & L_{a_2 0_{3\oplus \bar{1}}}                 \\
\\[-6pt]\hline\\[-6pt]
large non-BPS $z_H \neq 0$            & \tyoung(-+-,+-+,-,+)                                                                                                                        & \frac{\SO_0(4,4)}{\SO(1,1,\field{R})\times\SO(1,1,\field{R})\ltimes[(\rep{(2,2)\oplus(3,1)})^{(2)}\oplus{\rep 1}^{(4)}]} & 18                                & \frac{[\SL(2,\field{C})]^4}{\field{C}^3}                                                                 & \tfrac{i}{\sqrt{2}}(\ket{0001}+\ket{0010}-\ket{0111}-\ket{1011})& $\in$  & L_{ab_3}                                    \\
\\[-6pt]\hline\\[-6pt]
                                      & \tyoung(-+-+-,+,-,+)                                                                                                                        & \frac{\SO_0(4,4)}{\SO(2,1;\field{R})\ltimes[\rep{1}^{(2)}\oplus\rep{3}^{(4)}\oplus\rep{1}^{(6)}]}                        &                                   &                                                                                                          &                                                                 &        &                                             \\
\\[-6pt]
``extremal''                          & \tyoung(+-+-+,-,+,-)                                                                                                                        & \frac{\SO_0(4,4)}{\SO(1,2;\field{R})\ltimes[\rep{1}^{(2)}\oplus\rep{3}^{(4)}\oplus\rep{1}^{(6)}]}                        & 20                                & \frac{[\SL(2,\field{C})]^4}{\SO(2,\field{C})\times\field{C}}                                             & i\ket{0001}+\ket{0110}-i\ket{1011}                              & $\in$  & L_{a_4}                                     \\
\\[-6pt]
                                      & \begin{pmatrix}I,II&\tyoung(+-+-,-+-+)& I,II\end{pmatrix}                                                                                   & \frac{\SO_0(4,4)}{\Sp(2,\field{R})\ltimes[\rep{1}^{(2)}\oplus\rep{3}^{(4)}\oplus\rep{1}^{(6)} ]}                         &                                   &                                                                                                          &                                                                 &        &                                             \\
\\[-6pt]\hline\hline\\[-6pt]
large \half BPS and non-BPS  $z_H=0$  & \begin{array}{c}\begin{pmatrix}I,II&\tyoung(+-+,+-+,-,-)\end{pmatrix}\\\begin{pmatrix}\;\tyoung(-+-,-+-,+,+)&I,II\end{pmatrix}\end{array}   & \frac{\SO_0(4,4)}{\SO(2,\field{R})\times\SO(2,\field{R})\ltimes[(\rep{(2,2)\oplus(3,1)})^{(2)}\oplus{\rep 1}^{(4)}]}     & 18                                & \frac{[\SL(2,\field{C})]^4}{[\SO(2,\field{C})]^2\times \field{C}}                                        & \ket{0000}+\ket{0111}                                           & $\in$  & L_{0_{3\oplus \bar{1}} 0_{3\oplus \bar{1}}} \\
\\[-6pt]\hline\\[-6pt]
``extremal''                          & \begin{array}{c}\begin{pmatrix}I,II&\tyoung(-+-+-,+-+)\end{pmatrix}\\\begin{pmatrix}\;\tyoung(+-+-+,-+-)&I,II\end{pmatrix}\end{array}       & \frac{\SO_0(4,4)}{\field{R}^{3(2)}\oplus\field{R}^{1(4)}\oplus\field{R}^{2(6)}}                                          & 22                                & \frac{[\SL(2,\field{C})]^4}{\field{C}}                                                                   & \ket{0000}+\ket{0101}+\ket{1000}+\ket{1110}                     & $\in$  & L_{0_{5\oplus \bar{3}}}                     \\
\\[-6pt]\hline\\[-6pt]
``extremal''                          & \begin{array}{c}\begin{pmatrix}I,II&\tyoung(+-+-+-+,-)\end{pmatrix}\\\begin{pmatrix}\;\tyoung(-+-+-+-,+)&I,II\end{pmatrix}\end{array}       & \frac{\SO_0(4,4)}{\field{R}^{(2)}\oplus\field{R}^{2(6)}\oplus\field{R}^{(10)}}                                           & 24                                & \frac{[\SL(2,\field{C})]^4}{\ \mathds{I}}                                                                         & \ket{0000}+\ket{1011}+\ket{1101}+\ket{1110}                     & $\in$  & L_{0_{7\oplus \bar{1}}}                     \\
\hline\hline
\end{tabular*}
\end{sidewaystable}
\endgroup

\subsection{\label{Large-Extremal}``Large'' (\textit{i.e.} Attractor) Extremal $%
D=4$\ $STU$\ Black Holes}

The five nilpotent orbits of $\SO_{0}\left( 4,4\right) $ of dim$_{\mathds{R}%
}=18$ \cite{Collingwood:1993} (which correspond to $L_{0_{3\oplus \overline{1}%
}0_{3\oplus \overline{1}}}$ and $L_{ab_{3}}$ four qubits
entanglement families \cite{Borsten:2010db}) are related to extremal
``large'' (and thus attractor) $STU$ $D=4$ black holes (BHs). As
discussed \textit{e.g}. in App. A.3 of \cite{Bossard:2009we}, they
are characterized by
\begin{eqnarray}
A^{5} &=&0;  \label{1} \\
V^{3} &=&S^{3}=C^{3}=0,  \label{2}
\end{eqnarray}
where $A\equiv \mathbf{28}$, $S\equiv \mathbf{8}_{s}$, $V\equiv \mathbf{8}%
_{v}$ and $C\equiv \mathbf{8}_{c}$ respectively denote the adjoint, vector,
spinor and conjugate spinor irreprs. of the $D=3$ U-duality group $%
G_{3,STU}=\SO_{0}\left( 4,4\right) $. Thus, the \textit{triality
}symmetry exhibited by the $\mathcal{N}=2$, $D=4$ $STU$ model
\cite{Duff:1995sm,Behrndt:1996hu} can be traced back to the triality
of irreprs. $\mathbf{8}$'s of $G_{3,STU}$
itself\footnote{%
In general, the relevant non-compact subalgebra $\frak{h}_{3}^{\ast }=\frak{g%
}_{4}$ for the application of the Kostant-Sekiguchi Theorem (\cite
{Collingwood:1993}, and Refs. therein) to the issue of extremal BHs is the
unique non-compact form of $\frak{h}_{3}$ ($\frak{h}_{3}\oplus \frak{su}%
\left( 2\right) $ being the maximal compact subalgebra symmetrically
embedded into $\frak{g}_{3}$) such that it is embedded \textit{maximally}
(through a commuting $\frak{sl}\left( 2,\mathds{R}\right) $ algebra) and
\textit{symmetrically} into $\frak{g}_{3}$ itself. At geometric level, $%
\frak{h}_{3}^{\ast }$ is selected through the $c^{\ast }$map, which
is the generalization, pertaining to timelike $D=4\rightarrow D=3$
reduction, of
the $c$-map \cite{Cecotti:1988qn} (for a review, and a list of Refs., see \textit{e.g.%
} \cite{Bergshoeff:2008be}). Thus, in the $STU$ case ($\frak{g}_{3,STU}=\frak{so}%
\left( 4,4\right) $, $\frak{h}_{3,STU}^{\ast }=\frak{sl}\left( 2,\mathds{R}%
\right) \oplus \frak{sl}\left( 2,\mathds{R}\right) \oplus \frak{sl}\left( 2,%
\mathds{R}\right) $), the ``black hole/qubit correspondence''
\cite{Borsten:2010db} exploited through the Kostant-Sekiguchi
Theorem (for identity connected components), enjoys a geometrical
interpretation in terms of $c^{\ast }$-map (\textit{e.g.} see
explicit treatment of $c^{\ast }$-map of $STU$ model in
\cite{Bergshoeff:2008be,Chemissany:2009hq}).}. Conditions (\ref{1})
and (\ref{2}) are exactly the ones requested for extremal ``large''
(and thus attractor) $STU$ $D=4$ BHs (see Eq. (\ref{nihil-3})
below).

For use in the subsequent treatment, let us introduce the following maps of
cyclical index permutations: the \textit{triality} $\tau $ (pertaining to $%
D=4$; $\mathds{I}$ denotes the identity throughout)
\begin{equation}
\tau :2\longrightarrow 3\longrightarrow 4\longrightarrow 2;~\tau ^{3}=%
\mathds{I};  \label{tau-def}
\end{equation}
and the \textit{quaterniality} $\pi $ (pertaining to $D=3$)
\begin{equation}
\pi :1\longrightarrow 2\longrightarrow 3\longrightarrow 4\longrightarrow
1;~\pi ^{4}=\mathds{I}.  \label{pi-def}
\end{equation}
As evident from the treatment given below, $\tau $ does commute with
$D=4$ supersymmetry, whereas $\pi $ does or does not, depending on
the case.

\subsubsection{\label{stu<--->N=8}$STU$ Parametrization of $\mathcal{N}=8$, $%
D=4 $ Supergravity}

The supergravity interpretation of the $\SO_{0}\left( 4,4\right)
$-nilpotent orbits of dim$_{\mathds{R}}\leqslant 18$ considered
below is based on the so-called ``$STU$ parametrization'' of
$\mathcal{N}=8$, $D=4$ supergravity,
discussed \textit{e.g.} in \cite{Ferrara:2006em}. This amounts to identifying the $%
\mathcal{N}=2$ central charge and the three $STU$ matter charges with the
four skew-eigenvalues $z_{i}$ ($i=1,...,4$ throughout) of the $\mathcal{N}=8$
central charge matrix as follows \cite{Ferrara:2006em,Ceresole:2009vp}
\begin{equation}
Z\equiv z_{1};~\sqrt{g^{s\overline{s}}}\overline{D}_{\overline{s}}\overline{Z%
}\equiv iz_{2};~\sqrt{g^{t\overline{t}}}\overline{D}_{\overline{t}}\overline{%
Z}\equiv iz_{3};~\sqrt{g^{u\overline{u}}}\overline{D}_{\overline{u}}%
\overline{Z}\equiv iz_{4}.  \label{stu-identifications}
\end{equation}
Thus, the effective BH potential $V_{BH}$, its criticality conditions (%
\textit{alias} the Attractor Eqs.) and the quartic invariant $\mathcal{I}%
_{4} $ of $\mathcal{N}=2$, $D=4$ $STU$ model can be traded for the
ones pertaining to maximal supergravity, respectively reading \cite
{Ferrara:2006em,Kallosh:1996uy,Ferrara:1997ci}:
\begin{eqnarray}
V_{BH} &=&\sum_{i}\left| z_{i}\right| ^{4};  \label{V_BH} \\
\partial _{\phi }V_{BH} &=&0\Leftrightarrow z_{i}z_{j}+\overline{z_{k}}%
\overline{z_{l}}=0,~\forall i\neq j\neq k\neq l;  \label{N=8-AEs-gen} \\
\mathcal{I}_{4} &=&\sum_{i}\left| z_{i}\right| ^{4}-2\sum_{i<j}\left|
z_{i}\right| ^{2}\left| z_{j}\right| ^{2}+4\left( \prod_{i}z_{i}+\prod_{i}%
\overline{z_{i}}\right) ,  \label{I4}
\end{eqnarray}
where the notation ``$i\neq j\neq k\neq l$'' means all different indices
throughout.

In the subsequent treatment, we will also consider $\mathcal{N}=4$,
$D=4$ supergravity, and we will use the ($\mathcal{N}=2$ $STU$
analogue of the) $\mathcal{N}=4$, $D=4$ normal frame adopted in
\cite{Andrianopoli:2010bj}, which is more convenient to unravel the
relations to $\mathcal{N}=8$, $D=4$ supergravity, and the
corresponding \textit{quaterniality} properties.

Due to maximal $\mathcal{N}=8$ supersymmetry, note that (\ref{V_BH}), (\ref
{N=8-AEs-gen}) and (\ref{I4}) are manifestly $\pi $-invariant, as it can be
checked at a glance by recalling (\ref{pi-def}). By performing a suitable $%
\mathcal{R}$-symmetry $\SU\left( 8\right) $-transformation, the
Hua-Bloch-Messiah-Zumino theorem \cite{Hua:1944,Zumino:1962,Bloch:1962} allows one to set the phases
of $z_{i}$'s to be all equal, namely:
\begin{equation}
z_{i}\equiv \left| z_{i}\right| e^{i\frac{\varphi }{4}},~\forall i,~\varphi
\in \left[ 0,8\pi \right) .
\end{equation}
This has been named ``special normal frame'' in \cite{Ferrara:2009bw}. It should
also be pointed out that, out of (\ref{N=8-AEs-gen}), only some of them are
independent up to $\pi $-transformations and complex conjugation, namely:
\begin{equation}
\left\{
\begin{array}{l}
z_{1}z_{2}+\overline{z_{3}}\overline{z_{4}}=0; \\
\\
z_{1}z_{3}+\overline{z_{2}}\overline{z_{4}}=0.
\end{array}
\right.  \label{N=8-AEs}
\end{equation}

\subsubsection{\label{Large-I4>0}$\left( A,B,C,D\right)$-GHZ Classes $%
\Leftrightarrow L_{0_{3\oplus \overline{1}}0_{3\oplus \overline{1}}}$ : $%
\frac{1}{2}$-BPS and Non-BPS $Z_{H}=0$ ``Large'' BHs}

The four nilpotent orbits corresponding to the family $L_{0_{3\oplus
\overline{1}}0_{3\oplus \overline{1}}}$ are classes of bi-separable four
qubit entanglement, namely $A$-GHZ, $B$-GHZ, $C$-GHZ and $D$-GHZ \cite{Borsten:2010db}%
. The corresponding Young tableaux are related through $\pi $, and they
actually reduce to only one up to $\pi $-transformations.

It is convenient to set $i\in \mathds{N}$ \textit{mod.} $4$ (\textit{i.e.} $%
i+4\equiv i$). Then, $\mathcal{N}=8$, $D=4$ $\frac{1}{8}$-BPS solutions to (%
\ref{N=8-AEs}) are given by \cite{Ferrara:2006em}
\begin{equation}
\forall i\left\{
\begin{array}{l}
z_{i}\neq 0; \\
\\
z_{i+1}=z_{i+2}=z_{i+3}=0,
\end{array}
\right.   \label{N=8-1/8-BPS-Attr}
\end{equation}
with $\varphi $ \textit{undetermined} (thus, the non-vanishing $z_{i}$'s are
generally complex). It is evident that the four solutions (\ref
{N=8-1/8-BPS-Attr}) are related through $\pi $. They exhibit the maximal
compact symmetry consistent with the charge orbit \cite{Ferrara:1997uz,Lu:1997bg,Borsten:2010aa,Cerchiai:2009pi}
(see also \cite{Ferrara:2007tu} for a treatment of ``moduli spaces'' of
attractors)
\begin{equation}
\mathcal{O}_{\mathcal{N}=8,\frac{1}{8}-BPS,\text{large}}=\frac{E_{7\left(
7\right) }}{E_{6(2)}},
\end{equation}
namely $\SU\left( 2\right) \times \SU(6)$ (obtained through the symmetry
enhancement at the BH event horizon), which is the \textit{maximal compact
subgroup} (\textit{mcs}) of the stabilizer $E_{6(2)}$.

In the $STU$ model, lower $\mathcal{N}=2$ supersymmetry puts one of the four
$\mathcal{N}=8$ skew-eigenvalues, say $z_{1}$ (without loss of generality,
up to re-labelling), on a \textit{primus inter pares} status, corresponding
to $\mathcal{N}=2$ central charge. Thus, solutions (\ref{N=8-1/8-BPS-Attr})
split into ($a=2,3,4$ throughout)
\begin{eqnarray}
\frac{1}{2}\text{-BPS} &:&\left\{
\begin{array}{l}
z_{1}=0; \\
\\
~z_{a}=0~\forall a;
\end{array}
\right.  \label{1/2-BPS} \\
&&  \notag \\
\left. \text{nBPS~}Z_{H}=0\right| _{a} &:&\left\{
\begin{array}{l}
z_{1}=0; \\
\\
z_{a}\neq 0; \\
\\
z_{b}=0,~\forall b\neq a~.
\end{array}
\right.  \label{nBPS-Z=0}
\end{eqnarray}
Note that the $STU$ \textit{triality} symmetry is implemented through $\tau $
(recall definition (\ref{tau-def})). Thus, $\mathcal{N}=8$ $\frac{1}{8}$-BPS
attractor solution (\ref{N=8-1/8-BPS-Attr}) splits into:

\begin{itemize}
\item  one $\tau $-invariant (\textit{i.e.} \textit{triality}-invariant) $%
\mathcal{N}=2$ $\frac{1}{2}$-BPS attractor solution (\ref{1/2-BPS});

\item  three $\mathcal{N}=2$ $STU$ non-BPS $Z_{H}=0$ solutions, related
through $\tau $ \cite{Bellucci:2008sv}. After the analysis in App. AII of \cite
{Bellucci:2006xz}, the solutions with $a=1$ and $a=\left\{ 2,3\right\} $ would
respectively correspond to class $II$ and class $I$ of non-BPS $Z_{H}=0$
attractors. However, in the $STU$ the corresponding charge orbits are
\textit{isomorphic }\cite{Bellucci:2006xz}, because of the underlying \textit{triality}
symmetry, \textit{cfr.} Eq. (\ref{iso-1}) below (see also \cite{Bellucci:2007zi}).
\end{itemize}

Furthermore, solutions (\ref{1/2-BPS})-(\ref{nBPS-Z=0}) have
different uplift properties to $\mathcal{N}=4$, $D=4$ supergravity
(with $n_{V}=6$ matter vector multiplets). In fact:

\begin{itemize}
\item  (\ref{1/2-BPS}) and (\ref{nBPS-Z=0}) with $a=1$ uplift to $\mathcal{N}%
=4$ $\frac{1}{4}$-BPS attractors;

\item  (\ref{nBPS-Z=0}) with $a=2,3$ uplift to non-BPS attractors with
vanishing horizon central charge ($\left. Z_{AB}\right| _{H}=0$)
\cite {Andrianopoli:2006ub}.
\end{itemize}

The resulting supersymmetry reduction scheme reads
\begin{equation}
\begin{array}{ccccc}
\underline{\mathcal{N}=8}: &  & \mathcal{O}_{\mathcal{N}=8,\frac{1}{8}-BPS,%
\text{large}} &  &  \\
&  & \downarrow & \searrow &  \\
\underline{\mathcal{N}=4}: &  & \mathcal{O}_{\mathcal{N}=4,n_{V}=6,\frac{1}{4%
}-BPS,\text{large}} &  & \mathcal{O}_{\mathcal{N}=4,n_{V}=6,nBPS,Z_{AB,H}=0,%
\text{large}} \\
&  & \SL\left( 2,\mathds{R}\right) \times \frac{\SO\left( 6,6\right) }{%
\SO\left( 2\right) \times \SO\left( 4,6\right) } &  & \SL\left( 2,\mathds{R}%
\right) \times \frac{\SO\left( 6,6\right) }{\SO\left( 2\right) \times \SO\left(
6,4\right) } \\
&  & \downarrow &  & \downarrow \\
\underline{\mathcal{N}=2}: &  &
\begin{array}{c}
\mathcal{O}_{\mathcal{N}=2,STU,\frac{1}{2}-BPS,\text{large}} \\
\updownarrow ^{\ast } \\
\mathcal{O}_{\mathcal{N}=2,STU,nBPS,Z_{H}=0,II,\text{large}}
\end{array}
&  & \mathcal{O}_{\mathcal{N}=2,STU,nBPS,Z_{H}=0,I,\text{large}}.
\end{array}
\label{d-1}
\end{equation}
``$\updownarrow ^{\ast }$'' indicates that the two orbits are related by the
exchange $z_{1}\longleftrightarrow z_{2}$, and \cite{Bellucci:2006xz} (see also \cite
{Borsten:2009yb})
\begin{equation}
\mathcal{O}_{\mathcal{N}=2,STU,\frac{1}{2}-BPS,\text{large}}\sim \mathcal{O}%
_{\mathcal{N}=2,STU,nBPS,Z_{H}=0,II,\text{large}}\sim \mathcal{O}_{\mathcal{N%
}=2,STU,nBPS,Z_{H}=0,I,\text{large}}=\frac{\left[ \SL\left( 2,\mathds{R}%
\right) \right] ^{3}}{\left[ \Un\left( 1\right) \right] ^{2}}.  \label{iso-1}
\end{equation}
Notice that the exchange $z_{1}\longleftrightarrow z_{2}$ of the two $%
\mathcal{N}=4$ skew-eigenvalues implies a flip of the sign of the $\mathcal{N%
}=2$ $H_{STU}=\left[ \Un\left( 1\right) \right] ^{3}$-invariant
function
\begin{equation}
\left| Z\right| ^{2}-g^{s\overline{s}}\left| D_{s}Z\right| ^{2},
\end{equation}
which in general allows one to discriminate between $\frac{1}{2}$-BPS
attractor and non-BPS $Z_{H}=0$ of class $II$ in the sequence of symmetric
special K\"{a}hler geometries based on $\mathds{R}\oplus \mathbf{\Gamma }%
_{1,n-1}$ (whose the $STU$ model is the $n=2$ element); see the discussion
in App. AII of \cite{Bellucci:2006xz}.

(\ref{d-1})-(\ref{iso-1}) correspond to the following chains of maximal
symmetric embeddings\footnote{%
Unless otherwise noted, in the present investigation all group embeddings
are maximal and symmetric.}, respectively for the numerator and the
stabilizer groups of the cosets:
\begin{eqnarray}
E_{7\left( 7\right) } &\supsetneq &\SL\left( 2,\mathds{R}\right) \times
\SO\left( 6,6\right) \supsetneq \left[ \SL\left( 2,\mathds{R}\right) \right]
^{3}\times \SO\left( 4,4\right) ;  \label{ch-1} \\
E_{6\left( 2\right) } &\supsetneq &\SO\left( 2\right) \times \SO\left(
6,4\right) \supsetneq \left[ \SO\left( 2\right) \right] ^{2}\times \SO\left(
4,4\right) .  \label{ch-1-bis}
\end{eqnarray}
Note that the groups of the chain (\ref{ch-1-bis}) are maximally and
symmetrically embedded into the group of the chain (\ref{ch-1}) only through
a factor group $\SO\left( 2\right) $.

As mentioned above, the correspondence of $\mathcal{N}=4$
supergravity with the maximal theory is highlighted within the
``democratic normal frame'' recently introduced in
\cite{Andrianopoli:2010bj}, in which the $\mathcal{R}$-symmetry
reduction in $D=4$:
\begin{equation}
\begin{array}{ccc}
\mathcal{N}=8 & \longrightarrow & \mathcal{N}=4,~n_{V}=6 \\
\SU\left( 8\right) & \supsetneq & \Un\left( 4\right) \times \SO(6)\sim \Un\left(
4\right) \times \SU\left( 4\right)
\end{array}
\end{equation}
is fully manifest. In such a ``democratic'' normal frame, the two
skew-eigenvalues $z_{1}$ and $z_{2}$ of the $\mathcal{N}=4$ central charge
are taken to be real non-negative through a suitable $\Un\left( 4\right) $%
-transformation, whereas the overall $\mathcal{N}=8$ phase $\varphi $
becomes, after a suitable $\SO(6)$-transformation, the overall phase in front
of the unique two non-vanishing components $\widetilde{\rho }_{1}\equiv
\left| z_{3}\right| $ and $\widetilde{\rho }_{2}\equiv \left| z_{4}\right| $
of $\mathcal{N}=4$ matter charges' vector $Z_{I}$ ($I=1,...,n_{V}=6$) \cite
{Andrianopoli:2010bj}.

As mentioned, from the QIT perspective \cite{Borsten:2010db} $\mathcal{N}=2$
supersymmetry singles out one qubit, say $A$(lice), on a \textit{primus
inter pares} status, because it is the (complexification of the) Ehlers $%
\SL\left( 2,\mathds{R}\right) $ determined by timelike
$D=4\longrightarrow D=3 $ reduction (see \textit{e.g.}
\cite{Bergshoeff:2008be} for a recent treatment and list of Refs.).
Thus, solution (\ref{1/2-BPS}) corresponds to an $A$-GHZ state,
whereas solutions (\ref{nBPS-Z=0}) correspond to $B$-GHZ, $C$-GHZ
and $D$-GHZ states \cite{Levay:2010ua}. This is also consistent with
the analysis of \cite {Chterental:2007}, characterizing
$L_{0_{3\oplus \overline{1}}0_{3\oplus \overline{1}}}$ as a
distinguished family of bi-separable four qubit states.

The $\mathcal{N}=2$ $D=4$ $STU$ interpretation given above is
further confirmed by the fact that the non-translational part of the
stabilizer of
the nilpotent $\SO_{0}\left( 4,4\right) $-orbits under consideration, \textit{%
i.e.} $\left[ \SO\left( 2;\mathds{R}\right) \right] ^{2}\sim \left[ \Un\left(
1\right) \right] ^{2}$, coincides\footnote{%
For a discussion of the relation between the nilpotent orbits of the $D=3$ $%
U $-duality group $G_{3}$ and the charge orbits of the corresponding $D=4U$%
-duality group $G_{4}$, see the end of Sec. 2.4 of \cite{Bossard:2009we}.} with the
stabilizer of the rank-$4$ GHZ orbits (see \cite{Bellucci:2006xz}, as well as Table VI
of \cite{Borsten:2009yb}) (\ref{iso-1}).

The fact that the solutions (\ref{N=8-1/8-BPS-Attr}) are related through $%
\pi $ corresponds to four Young tableaux which are related through
$D=3$
permutation symmetry. Correspondingly, there exist four nilpotent $%
\SO_{0}\left( 4,4\right) $-orbits of dimension $18$ (related to $%
L_{0_{3\oplus \overline{1}}0_{3\oplus \overline{1}}}$), which reduce to only
one up to $\pi $-transformations. This is consistent with the fact that the $%
\frac{1}{2}$-BPS ``large'' and both types ($I$ and $II$) of non-BPS
$Z_{H}=0$ charge orbits of $\mathcal{N}=2$, $D=4$ $STU$ model are
\textit{isomorphic}, as given by Eq. (\ref{iso-1})
\cite{Bellucci:2006xz}.

\subsubsection{\label{Large-I4<0}$4$-way Entanglement in $L_{ab_{3}}$ : Non-BPS
$Z_{H}\neq 0$ ``Large'' BHs}

The $\SO_{0}\left( 4,4\right) $-nilpotent orbit corresponding to the $%
L_{ab_{3}}$ four qubit entanglement family is related to a Young
tableaux which is invariant under $D=3$ permutations
\cite{Borsten:2010db}. Consistently, the corresponding solution to
(\ref{N=8-AEs}) is $\pi $-invariant \cite{Ferrara:2006em}:
\begin{equation}
\left\{
\begin{array}{l}
z_{i}=\rho e^{i\frac{\varphi }{4}},~\rho \in \mathds{R}_{0}^{+},~\forall i,
\\
\\
\varphi =\pi +2k\pi ,~k\in \mathds{Z}.
\end{array}
\right.  \label{nBPS}
\end{equation}
In $D=4$, solution (\ref{nBPS}) is non-BPS in $\mathcal{N}=8$, non-BPS with $%
\left. Z_{AB}\right| _{H}\neq 0$ ($A$, $B=1,...,4$) in $\mathcal{N}=4$ (with
$n_{V}=6$ matter vector multiplets), and non-BPS $Z_{H}\neq 0$ in $\mathcal{N%
}=2$ $STU$ model. It exhibits the maximal compact symmetry consistent with
\cite{Ferrara:1997uz,Lu:1997bg}
\begin{equation}
\mathcal{O}_{\mathcal{N}=8,nBPS}=\frac{E_{7\left( 7\right) }}{E_{6\left(
6\right) }},
\end{equation}
namely $\USp\left( 8\right) =mcs\left( E_{6\left( 6\right) }\right) $
(obtained through the symmetry enhancement at the black hole event horizon).

As mentioned, solutions (\ref{nBPS}) uplift to non-BPS attractors with $%
\left. Z_{AB}\right| _{H}\neq 0$ of $\mathcal{N}=4$, $D=4$,
$n_{V}=6$ supergravity \cite{Andrianopoli:2006ub}. The resulting
supersymmetry reduction pattern reads
\begin{equation}
\begin{array}{cc}
\underline{\mathcal{N}=8}: & \mathcal{O}_{\mathcal{N}=8,nBPS} \\
& \downarrow \\
\underline{\mathcal{N}=4}: &
\begin{array}{c}
\mathcal{O}_{\mathcal{N}=4,n_{V=6},nBPS,Z_{AB,H}\neq 0,\text{large}} \\
\\
\SL\left( 2,\mathds{R}\right) \times \frac{\SO\left( 6,6\right) }{\SO\left(
1,1\right) \times \SO\left( 5,5\right) }
\end{array}
\\
& \downarrow \\
\underline{\mathcal{N}=2}: &
\begin{array}{c}
\mathcal{O}_{\mathcal{N}=2,STU,nBPS,Z_{H}\neq 0,\text{large}} \\
\\
\frac{\left[ \SL\left( 2,\mathds{R}\right) \right] ^{3}}{\left[ \SO\left(
1,1\right) \right] ^{2}}.
\end{array}
\end{array}
\label{ostia-1}
\end{equation}
For the numerators of cosets (\ref{ch-1}) holds, whereas for the stabilizers
the following chain of embeddings holds:
\begin{equation}
E_{6\left( 6\right) }\supsetneq \SO\left( 1,1\right) \times \SO\left(
5,5\right) \supsetneq \left[ \SO\left( 1,1\right) \right] ^{2}\times \SO\left(
4,4\right) .
\end{equation}

In $\mathcal{N}=2$ $STU$ model, the manifest $\pi $-invariance of solution (%
\ref{nBPS}) corresponds to the fact that the central charge and matter
charges are set on the very same footing. This leads to the statement that
the corresponding four qubit state is four-way entangled \cite{Levay:2010ua}. In
turn, this is consistent with the analysis of \cite{Chterental:2007}, stating that all
families but $L_{0_{3\oplus \overline{1}}0_{3\oplus \overline{1}}}$ contain
four-way entangled states.

On the other hand, the $\mathcal{N}=2$ $STU$ interpretation given above is
confirmed by the fact that the non-translational part of the stabilizer of
this nilpotent $\SO_{0}\left( 4,4\right) $-orbit, \textit{i.e.} $\left[
\SO\left( 1,1;\mathds{R}\right) \right] ^{2}$, coincides with the stabilizer
of the rank-$4$ GHZ orbit (see \cite{Bellucci:2006xz}, as well as Table VI of \cite
{Borsten:2009yb}) given by (\ref{ostia-1}).

The fact that solution (\ref{nBPS}) is $\pi $-invariant corresponds
to a Young tableaux which is invariant under $D=3$ permutation
symmetry. Consistently, the corresponding nilpotent $\SO_{0}\left(
4,4\right) $-orbit of real dimension $16$ (related to $L_{ab_{3}}$)
is unique \cite{Borsten:2010db}, and
it maps to the non-BPS $Z_{H}\neq 0$ ``large'' charge orbit of $\mathcal{N}%
=2 $, $D=4$ $STU$ model \cite{Bellucci:2006xz}.

\subsubsection{\label{Invariants-Extremality}Conditions for Attractor
Extremality}

As discussed in Sect. VI of \cite{Levay:2010ua}, all orbits treated in Sect. \ref
{Large-Extremal} have all four $4$-qubit invariants vanishing. Namely, by
using the notations of \cite{Levay:2010ua} (see \textit{e.g.} App. VIII, as well as
Refs., therein):
\begin{equation}
I_{1}=I_{2}=I_{3}=I_{4}=0.  \label{3}
\end{equation}
The (permutation-invariant) quadratic $4$-qubit invariant $I_{1}$
has the physical interpretation of the extremality black hole
parameter $c^{2}$. Despite (\ref{3}), the various $4$-qubits states
still can be characterized through their entanglement properties.
Moreover, the orbits of Subsect. \ref {Large-I4>0} have
$\mathcal{I}_{4}>0$, whereas the orbit of Subsect. \ref {Large-I4<0}
has $\mathcal{I}_{4}<0$, with $\mathcal{I}_{4}$ denoting the quartic
($3$-qubits \cite{Borsten:2009yb}) $G_{4,STU}$-invariant.

Conditions (\ref{1})-(\ref{2}) are equivalent to (\ref{3}): they both%
\footnote{%
As to our knowledge, the only possible counter-example to this statement
might have been provided by the orbit $\mathcal{O}_{3K}^{\prime }$ of the $%
D=3$ timelike-reduced $S^{3}$ model, studied in \cite{Kim:2010bf}.
\par
Indeed, orbit $\mathcal{O}_{3K}^{\prime }$ can be obtained from a
``degeneration procedure'' (outlined \textit{e.g.} in Sect. 5 of \cite
{Bellucci:2007zi}, as well as at the very end of App. A.3 of \cite{Bossard:2009we}) of the
orbits discussed in Sect. \ref{Large-Extremal}. As given by Table 2 of \cite
{Kim:2010bf}, it may also have $\mathcal{I}_{4}=0$ (besides $\mathcal{I}%
_{4}>0$).
\par
However, just above the start of Subsect. 6.1.1 of
\cite{Kim:2010bf}, such an orbit is claimed to be
\textit{unphysical}, and thus to be disregarded.} are conditions to
have extremal ($c^{2}=0$) ``large'' ($\mathcal{I}_{4}\neq 0 $) $STU$
$D=4$ BHs, thus exhibiting an \textit{Attractor Mechanism} (\cite
{Ferrara:1995ih,Strominger:1996kf,Ferrara:1996dd,Ferrara:1996um,Ferrara:1997tw};
see Sect. VI of \cite{Levay:2010ua}, as well).

As summarized in Subsect. 1.4 of \cite{Bossard:2009we} (which in turn recalls the
treatment of \cite{Bossard:2009at}), the above conditions can be reformulated in a $%
D=3$ language as follows: any static, spherically symmetric,
asymptotically flat, extremal ``large'' black hole solution in $D=4$
theories with symmetric scalar manifold is characterized by
\begin{equation}
\left[ \left. Q\right| _{\mathbf{R}}\right] ^{3}=0,  \label{nihil-3}
\end{equation}
where $Q$ is the $\frak{g}_{3}$-valued $D=3$ Noether charge, and
$\mathbf{R}$ is the relevant irrepr. of $G_{3}$ (for instance, the
spinor $\mathbf{8}_{s}$ for $G_{3,STU}$). Among simple $D=3$
U-duality groups related to symmetric
scalar manifolds, the unique exception to nilpotency condition (\ref{nihil-3}%
) is provided by $G_{3}=E_{8\left( 8\right) }$ (maximal supergravity,
related to $J_{3}^{\mathds{O}_{s}}$), for which (\ref{nihil-3}) gets
replaced by
\begin{equation}
\left[ \left. Q\right| _{\mathbf{3875}}\right] ^{5}=0.  \label{nihil-5}
\end{equation}

It is also worth mentioning that in general, the condition
(\ref{nihil-3}) (or (\ref{nihil-5})) on $Q$ must be supplemented by
a condition on the conjugate $D=3$ geodesic $\left(
\frak{g}_{3}\ominus \frak{h}_{3}^{\ast }\right) $-valued momentum
(\textit{cfr.} definition (1.15) of \cite{Bossard:2009we})
\begin{equation}
P\equiv \mathcal{V}Q\mathcal{V}^{-1},
\end{equation}
where $\mathcal{V}$ is the $D=3$ coset representative. Such a condition on $%
P $ is discussed at the end of App. A.1 of \cite{Bossard:2009we}, and it has recently
been checked also in the $D=3$ timelike reduction of the so-called $\mathcal{%
N}=2$, $D=4$ $t^{3}$ model in \cite{Kim:2010bf}\footnote{%
Namely, one obtains the coincidence of $\beta $-label and $\gamma $-label in
physical solutions out of $\SO_{0}\left( 2,2\right) $-orbits $\mathcal{O}%
_{3K} $ and $\mathcal{O}_{4K}^{\prime }$; see Subsect. 4.4 and Table 2 of
\cite{Kim:2010bf}.}.

Moreover, it should be remarked that the treatment leading to nilpotency
condition (\ref{nihil-3}) (or (\ref{nihil-5})) is based on the assumption
made in \cite{Bossard:2009at}, that the extremal BHs under consideration can be
obtained through a limit procedure from non-extremal black hole solutions.
This observation will be reconsidered further below.

As treated in Sects. 5 and 6 of \cite{Bossard:2009at}, discussed at the end of
Subsect. 1.4 of \cite{Bossard:2009we}, and stated at the end of Subsect. 2.4 of \cite
{Bossard:2009we} itself, for all nilpotent $G_{3}$-orbits satisfying the condition (%
\ref{nihil-3}) (or (\ref{nihil-5}))\textit{\ the non-translational
part of the stabilizer coincides (at the horizon) with the
stabilizer of the corresponding ``large'' orbit of the relevant
}$D=4$\textit{\ charge irrepr.
of }$G_{4}$. For $\mathcal{N}=2$ ``magic'' octonionic ($J_{3}^{\mathds{O}%
_{s}}$) and $\mathcal{N}=8$ ($J_{3}^{\mathds{O}}$) supergravity
$D=4$
theories\footnote{%
This matching can be explicitly checked for
\par
\begin{enumerate}
\item  $J_{3}^{\mathds{O}_{s}}$, by making use of the results reported in
App. A.2 of \cite{Bossard:2009we} or equivalently, at the level of nilpotent $%
H_{3}^{\ast }\sim Spin^{\ast }\left( 16\right) $-strata, by considering Eqs.
(5.17) and (5.18) of \cite{Bossard:2009at};
\par
\item  $J_{3}^{\mathds{H}}$, pertaining to the ``twin'' \cite
{Andrianopoli:1996ve,Ferrara:2008ap,Roest:2009sn} theories $\mathcal{N}=6$
and $\mathcal{N}=2$ ``magic'' symmetric quaternionic. This case is
considered in Subsect. 3.1 of \cite{Bossard:2009we}, and the matching of orbit
stabilizers can be checked, equivalently at the level of nilpotent $%
H_{3}^{\ast }\sim \left( Spin^{\ast }\left( 12\right) \times \SL\left( 2,%
\mathds{R}\right) \right) $-strata, from Eqs. (5.13)-(5.15) of \cite{Bossard:2009at}.
\end{enumerate}
}, this is expressed by Eqs. (2.100) and (2.101) of \cite{Bossard:2009we}.

\subsection{\label{Small-Extremal}``Small'' (\textit{i.e.} Non-Attractor)
Extremal $D=4$\ $STU$\ BHs}

The theory of $G_{4}$-orbits in the relevant (real, symplectic) charge
representation space is known only for extremal\footnote{%
Let us point out that in (\cite{Bossard:2009at} and) \cite{Bossard:2009we} ``extremal'' is
used as synonym of ``large with $c^{2}=0$'', whereas in \cite{Levay:2010ua} and in
\cite{Kim:2010bf} ``extremal'' is used simply as synonym of ``$c^{2}=0$'',
and we will adopt this latter use. In fact, note that in \cite{Kim:2010bf}
various nilpotent $\widetilde{K}\equiv H_{3,t^{3}}^{\ast }=\SO_{0}\left(
2,2\right) $-orbits correspond to BHs with $c^{2}=0$ and $\mathcal{I}_{4}=0$%
, thus to ``small'' extremal BHs.} ``large'' and ``small'' $D=4$ BHs
in
theories with symmetric scalar manifolds $\frac{G_{4}}{H_{4}}$, where $%
H_{4}=mcs\left( G_{4}\right) $. In the supergravity theories related to

\begin{itemize}
\item  $J_{3}^{\mathds{O}_{s}}$ (see App. A.2 of \cite{Bossard:2009we}, as well as
Eqs. (5.17) and (5.18) of \cite{Bossard:2009at});

\item  $J_{3}^{\mathds{H}}$ (see Eqs. (5.13)-(5.15) of \cite{Bossard:2009at}),
\end{itemize}

the nilpotent $G_{3}$-orbits (or, equivalently, their relevant Lagrangian
submanifolds given by the corresponding nilpotent $H_{3}^{\ast }$%
-orbits/strata) related to extremal ``small'' $D=4$ BHs have real
dimension \textit{smaller} than the ones related to extremal
``large'' $D=4$ BHs, \textit{i.e.} than the ones satisfying
condition (\ref{nihil-3}) (or (\ref {nihil-5})).

Furthermore, in the aforementioned cases the $mcs$ of the
non-translational part of the stabilizer of each of these nilpotent
$G_{3}$-orbits can be checked to match the $mcs$ of the
non-translational part of the stabilizer of the corresponding
``small'' $G_{4}$-orbit, \textit{i.e.} the stabilizer of the
corresponding \textit{moduli space} (if any) \textit{of }$D=4$
\textit{ADM mass}. The corresponding $D=4$ BHs are ``small'' and
extremal, and therefore they all have $\mathcal{I}_{4}=0$ and
$c^{2}=0$. This latter relation implies $I_{1}=0$, where $I_{1}$ is
the quadratic $4$-qubit invariant (see \textit{e.g.}
\cite{Levay:2010ua} and Refs. therein).

In the $\mathcal{N}=2$ $STU$ model, the situation is rather peculiar,
because the groups involved are small, and they may also lack a non-trivial $%
mcs$. Actually, the three non-translational part of the stabilizers of rank-$%
3$ (\textit{lightlike}), $2$ (\textit{critical}), $1$ (\textit{%
doubly-critical}) orbits of Table VI of \cite{Borsten:2009yb} respectively
are: $\mathds{I}$, $\SpO\left( 2,1\right) $ and $\left[ \SO\left( 1,1\right) %
\right] ^{2}$, with $mcs$ given by $\mathds{I}$, $\SO\left( 2\right) $ and $%
\mathds{I}$, respectively.\medskip

All this leads to the following statement: within the considered framework,
the $G_{3}$-nilpotent orbits with dimension \textit{smaller} (corresponding
to a nilpotency degree \textit{lower}) than the one of the $G_{3}$-nilpotent
orbits satisfying condition (\ref{nihil-3}) (or (\ref{nihil-5})) correspond
to ``small'' extremal $D=4$ BHs. As a consequence, the sets of $G_{3}$%
-nilpotent orbits (grouped under $D=3$ permutation symmetry) are in
one-to-one correspondence with the classes of ``small'' charge orbits of $%
G_{4}$.\smallskip\

Within the $\mathcal{N}=2$ $STU$ model, we are now going to work out
in detail the correspondence among the rank-$3$, $2$, $1$ orbits of
Table VI of \cite{Borsten:2009yb} and the (classes of) nilpotent
$\SO_{0}\left( 4,4\right) $-orbits of real dimension $16$, $12$ and
$10$, outlined in \cite {Borsten:2010db}. The $STU$ model turns out
to exhibit an high degree of ``degeneration'': the BPS and non-BPS
$3$-charge orbits all are isomorphic, as well as all $2$-charge
orbits are. Furthermore, the $D=3$ permutation properties of the
related Young tableaux can be inferred from the action of cyclic
permutations $\pi $ on the representative solutions of the
corresponding $G_{4,STU}=\left[ \SL\left( 2,\mathds{R}\right) \right] ^{3}$%
-invariant constraints defining the ``small'' orbits of the $\left( \mathbf{2%
},\mathbf{2},\mathbf{2}\right) $ irrepr. of $G_{4,STU}$.

\subsubsection{$A$-$W$ Classes $\Leftrightarrow L_{a_{2}0_{3}\oplus \overline{1}}$
: \textit{Lightlike} $\frac{1}{2}$-BPS and non-BPS Orbits}

The $3$-charge (\textit{lightlike}) ``small'' orbit of the $\left( \mathbf{2}%
,\mathbf{2},\mathbf{2}\right) $ of $G_{4,STU}$ is given by (see
Table VI of \cite{Borsten:2009yb})
\begin{equation}
\mathcal{O}_{STU,lightlike}=\frac{\left[ \SL\left(
2,\mathds{R}\right) \right]
^{3}}{\mathds{R}^{2}},~\text{dim}_{\mathds{R}}=7.
\label{stu-3-charge}
\end{equation}
As given by the treatment below, such an orbit actually corresponds to three
isomorphic orbits, which for a generic element of the Jordan symmetric
sequence $\mathds{R}\oplus \mathbf{\Gamma }_{1,n_{V}-2}$ with $%
n_{V}\geqslant 4$, are distinct \cite{ICL-2}.

The $\left[ \SL\left( 2,\mathds{R}\right) \right] ^{3}$-invariant constraint
which defines $\mathcal{O}_{STU,lightlike}$ is simply the vanishing of $%
\mathcal{I}_{4}$:
\begin{equation}
\mathcal{I}_{4}=\sum_{i}\left| z_{i}\right| ^{4}-2\sum_{i<j}\left|
z_{i}\right| ^{2}\left| z_{j}\right| ^{2}+4\left( \prod_{i}z_{i}+\prod_{i}%
\overline{z_{i}}\right) =0.  \label{3-charge-constraint}
\end{equation}
As a consequence of Eq. (\ref{I4}), the constraint (\ref{3-charge-constraint}%
) is manifestly $\pi $-invariant.

A set of representative solutions to the constraint (\ref
{3-charge-constraint}) reads:
\begin{equation}
\left\{
\begin{array}{l}
\left| z_{i}\right| \equiv \mathcal{A}; \\
\\
\left| z_{i+1}\right| =\left| z_{i+2}\right| =\left| z_{i+3}\right| \equiv
\mathcal{B}\;\neq \mathcal{A},~
\end{array}
\right.  \label{3-charge-sols-mcs}
\end{equation}
with $\mathcal{A},\mathcal{B}\in \mathds{R}_{0}^{+}$, and
\begin{equation}
\mathcal{A}^{4}-3\mathcal{B}^{4}-6\mathcal{A}^{2}\mathcal{B}^{2}+8\mathcal{AB%
}^{3}\cos \varphi =0.  \label{3-charge-sols-mcs-2}
\end{equation}
The $z_{i}$'s are generally complex. The solutions (\ref{3-charge-sols-mcs}%
)-(\ref{3-charge-sols-mcs-2}) exhibit the maximal compact symmetry
consistent with \cite{Ferrara:1997uz,Lu:1997bg}
\begin{equation}
\mathcal{O}_{\mathcal{N}=8,\frac{1}{8}-BPS,\text{small}}=\frac{E_{7\left(
7\right) }}{F_{4\left( 4\right) }\rtimes \mathds{R}^{26}},
\end{equation}
namely $\SU\left( 2\right) \times \USp(6)=mcs\left( F_{4\left( 4\right)
}\right) $.

Another set of four representative solutions to the constraint (\ref
{3-charge-constraint}) is given by
\begin{equation}
\text{~}\left\{
\begin{array}{l}
z_{i}=0; \\
\\
\multicolumn{1}{c}{\left| z_{i+1}\right| ^{4}+\left| z_{i+2}\right|
^{4}+\left| z_{i+3}\right| ^{4}-2\left( \left| z_{i+1}\right| ^{2}\left|
z_{i+2}\right| ^{2}+\left| z_{i+1}\right| ^{2}\left| z_{i+3}\right|
^{2}+\left| z_{i+2}\right| ^{2}\left| z_{i+3}\right| ^{2}\right) =0.}
\end{array}
\right.  \label{3-charge-sols}
\end{equation}
\ The $z_{i}$'s of solutions (\ref{3-charge-sols}) are generally complex.
The solutions (\ref{3-charge-sols}) exhibit the generic compact symmetry
(consistent with the structure of the skew-diagonalized $\mathcal{N}=8$, $%
D=4 $ central charge matrix itself) $\left[ \SU\left( 2\right) \right] ^{4}$
(recall that $\USp\left( 2\right) =\SU(2)$).

In both sets (\ref{3-charge-sols-mcs})-(\ref{3-charge-sols-mcs-2})
and (\ref {3-charge-sols}), the four solutions are related through
the iterated action of $\pi $. This exactly corresponds to the $D=3$
permutation properties of the Young tableaux of the $\SO_{0}\left(
4,4\right) $-nilpotent orbits of real dimension $16$, in turn
related to the four $A$-$W$ classes belonging to the family
$L_{a_{2}0_{3}\oplus \overline{1}}$ \cite{Borsten:2010db}: such
Young tableaux are related through $D=3$ permutation symmetry, and
they reduce to a unique one, and thus to a unique $\SO_{0}\left(
4,4\right) $-nilpotent orbit, up to $D=3$ permutations.

A direct comparison of (\ref{3-charge-sols-mcs}) and (\ref{N=8-1/8-BPS-Attr}%
) explains the analogous transformation properties under $\pi $, as well as
the analogous structure of Young tableaux characterizing the set (\ref
{3-charge-sols-mcs}) of representative $3$-charge solutions and the set (\ref
{N=8-1/8-BPS-Attr}) of attractor solutions with $\mathcal{I}_{4}>0$ (also
recall that (\ref{3-charge-sols-mcs}) does not admit $\mathcal{AB}=0$).
Indeed, the limit $\mathcal{B}=0$ of Eq. (\ref{3-charge-sols-mcs}) leads to
Eq. (\ref{N=8-1/8-BPS-Attr}), and the corresponding (maximal) manifest
compact symmetry gets enhanced from $\SU(2)\times \USp(6)$ (pertaining to $%
\mathcal{O}_{\mathcal{N}=8,\frac{1}{8}-BPS,\text{small }}$) to $\SU(2)\times
\SU(6)$ (pertaining to $\mathcal{O}_{\mathcal{N}=8,\frac{1}{8}-BPS,\text{%
large }}$).\medskip

It is here worth remarking that, despite they share the same
supersymmetry properties in $\mathcal{N}=8$, $D=4$

supergravity, Eqs. (\ref{N=8-1/8-BPS-Attr}) and (\ref{3-charge-sols-mcs})
(or (\ref{3-charge-sols})) have different space-time properties, related to
the \textit{Attractor Mechanism }\cite{Ferrara:1995ih,Strominger:1996kf,Ferrara:1996dd,Ferrara:1996um,Ferrara:1997tw}:

\begin{itemize}
\item  Eq. (\ref{N=8-1/8-BPS-Attr}) has a space-time localization at the
black hole event horizon, where the symmetry enhancement $[\SU(2)]^{4}$ $%
\longrightarrow $ $\SU(2)\times \SU(6)$ due to \textit{Attractor Mechanism}
takes place. Furthermore, Eq. (\ref{N=8-1/8-BPS-Attr}) is a solution to the $%
\SU\left( 8\right) $-invariant Attractor Eqs., and it stabilizes the scalars
in terms of the charges, fixing one point in the scalar manifold (for a
given input of charges).

\item  Eq. (\ref{3-charge-sols-mcs}) holds all along the scalar flow (for
every value of the radial coordinate $r$), because, since the corresponding
extremal black hole is ``small'', there's no event horizon at which the
scalars should be stabilized. Furthermore, since Eq. (\ref{3-charge-sols-mcs}%
) is a complete set of representative solutions to the $G_{4}$-invariant
constraint $\mathcal{I}_{4}=0$, it does not stabilize the scalars in terms
of the charges, and thus it holds in all scalar manifold.
\end{itemize}

The difference in space-time localization and
``scalar-manifold-localization'' of Eqs. (\ref{N=8-1/8-BPS-Attr}) and (\ref
{3-charge-sols-mcs}) is originated by the interplay between the \textit{%
Attractor Mechanism} and the U-duality (\textit{i.e. }$G_{4}$%
-)invariance.\medskip

In the treatment below (refining the results of
\cite{Borsten:2010db}), we show how the four Young tableaux
associated to $L_{a_{2}0_{3}\oplus \overline{1}}$ can be related to
one $\frac{1}{2}$-BPS and three non-BPS $3$-charge representative
solutions of $STU$ model, related through the iterated action of
$\pi $. Such a $D=4$ supersymmetry interpretation can be summarized
by the following scheme \cite{ICL-2}:
\begin{equation}
\begin{array}{ccccc}
\underline{\mathcal{N}=8}: &  & \mathcal{O}_{\mathcal{N}=8,\frac{1}{8}-BPS,%
\text{small}} &  &  \\
&  & \downarrow & \searrow &  \\
\underline{\mathcal{N}=4}: &  & \mathcal{O}_{\mathcal{N}=4,n_{V=6},\mathbf{%
C1~}\left[ \frac{1}{4}-BPS\right] } &  & \mathcal{O}_{\mathcal{N}=4,n_{V=6},%
\mathbf{C2~}\left[ nBPS\right] } \\
&  & \SL\left( 2,\mathds{R}\right) \times \frac{\SO\left( 6,5\right) }{%
\SO\left( 4,5\right) \rtimes \left( \mathds{R}^{4,5}\times \mathds{R}\right) }
&  & \SL\left( 2,\mathds{R}\right) \times \frac{\SO\left( 6,5\right) }{%
\SO\left( 5,4\right) \rtimes \left( \mathds{R}^{5,4}\times \mathds{R}\right) }
\\
&  & \downarrow &  & \downarrow \\
\underline{\mathcal{N}=2}: &  &
\begin{array}{c}
\mathcal{O}_{\mathcal{N}=2,STU,\mathbf{C1~}\left[
\frac{1}{2}-BPS\right] }
\\
\updownarrow \ast \\
\mathcal{O}_{\mathcal{N}=2,STU,\mathbf{C1~}\left[ nBPS\right] }
\end{array}
&  & \mathcal{O}_{\mathcal{N}=2,STU,\mathbf{C2~}\left[ nBPS\right]
},
\end{array}
\label{ddd-1}
\end{equation}
where ``$\mathbf{C1}$'' and ``$\mathbf{C2}$'' refer to the classification of
\cite{Cerchiai:2009pi,Andrianopoli:2010bj,Ceresole:2010nm}, and it holds that
\begin{eqnarray}
\mathcal{O}_{\mathcal{N}=2,STU,\mathbf{C1~}\left[
\frac{1}{2}-BPS\right] } &\sim
&\mathcal{O}_{\mathcal{N}=2,STU,\mathbf{C1~}\left[ nBPS\right] }\sim
\mathcal{O}_{\mathcal{N}=2,STU,\mathbf{C2~}\left[ nBPS\right] }
\label{ddd-2} \\
&\sim &\mathcal{O}_{STU,lightlike}=\frac{\left[ \SL\left(
2,\mathds{R}\right) \right] ^{3}}{\mathds{R}^{2}},  \label{ddd-3}
\end{eqnarray}
consistent with Table VI of \cite{Borsten:2009yb}. Consequently, Eqs. (\ref
{ddd-1})-(\ref{ddd-3}) correspond to the following chains of embeddings: for
the numerators of cosets (\ref{ch-1}) holds, whereas for the stabilizers it
holds ($\mathds{R}\times \mathds{R\equiv R}^{2}$):
\begin{eqnarray}
F_{4\left( 4\right) }\rtimes \mathds{R}^{26} &\supsetneq &\SO\left(
5,4\right) \rtimes \left( \mathds{R}^{16}\times \mathds{R}^{5,4}\times
\mathds{R}\right)  \notag \\
&\supsetneq &\SO\left( 4,4\right) \rtimes \left( \mathds{R}^{8_{s}}\times
\mathds{R}^{8_{c}}\times \mathds{R}^{8_{v}}\times \mathds{R}^{2}\right) .
\label{ch-2}
\end{eqnarray}
In (\ref{ch-2}), the spinor $\mathds{R}^{16}$ and the \textit{triality}%
-symmetric product $\mathds{R}^{8_{s}}\times \mathds{R}^{8_{c}}\times
\mathds{R}^{8_{v}}$ are progressively truncated out.\medskip\

We now give a set of four independent representative solutions to lightlike
constraint (\ref{3-charge-constraint}). They are particular,
maximally-symmetric solutions of the type (\ref{3-charge-sols}). We will
explain their relation to the four Young tableaux of the four $\SO_{0}\left(
4,4\right) $-nilpotent orbits of real dimension $16$, in turn related to the
four $A$-$W$ classes belonging to the family $L_{a_{2}0_{3}\oplus \overline{1}%
} $ \cite{Borsten:2010db}.

As mentioned above, within the subsequent analysis, we will use the ($%
\mathcal{N}=2$ $STU$ analogue of the) $\mathcal{N}=4$ normal frame
adopted in \cite{Andrianopoli:2010bj}, in which the relation to
$\mathcal{N}=8$, $D=4$ supergravity (and the corresponding
\textit{quaterniality} properties) are more manifest.

\begin{enumerate}
\item  The first prototypical representative solution reads
\begin{equation}
\left[ \mathbf{1.i}\right] :\left\{
\begin{array}{l}
z_{1}=0; \\
\\
2\left| z_{3}\right| =2\left| z_{4}\right| =\left| z_{2}\right| .
\end{array}
\right.
\end{equation}
This is a non-BPS solution: it is the $z_{1}=0$ limit of the representative
solution of orbit $\mathcal{O}_{\mathcal{N}=2,STU,\mathbf{C1~}\left[ nBPS%
\right] }$, given by the ($\mathcal{N}=2$ $STU$ analogue of the) ``$%
z_{2}>z_{1}$ counterpart'' of the solution treated at point C1) of Subsect.
4.4 of \cite{Andrianopoli:2010bj}. By applying triality transformation $\tau $ defined
in (\ref{tau-def}), from $\left[ \mathbf{1.i}\right] $ one can generate
other two equivalent $\mathcal{N}=2$ non-BPS solutions, belonging to $%
\mathcal{O}_{\mathcal{N}=2,STU,\mathbf{C1~}\left[ nBPS\right] }$
itself, namely:
\begin{eqnarray}
\left[ \mathbf{1.ii}\right] &:&\left\{
\begin{array}{l}
z_{1}=0; \\
\\
2\left| z_{4}\right| =2\left| z_{2}\right| =\left| z_{3}\right| .
\end{array}
\right. \\
\left[ \mathbf{1.iii}\right] &:&\left\{
\begin{array}{l}
z_{1}=0; \\
\\
2\left| z_{2}\right| =2\left| z_{3}\right| =\left| z_{4}\right| .
\end{array}
\right.
\end{eqnarray}
Solutions $\left[ \mathbf{1.i}\right] $-$\left[ \mathbf{1.iii}\right] $ are $%
\tau $-equivalent, because it holds that:
\begin{equation}
\left[ \mathbf{1.i}\right] \overset{\tau }{\longrightarrow }\left[ \mathbf{%
1.ii}\right] \overset{\tau }{\longrightarrow }\left[ \mathbf{1.iii}\right]
\overset{\tau }{\longrightarrow }\left[ \mathbf{1.i}\right] .
\end{equation}
They all are $\mathcal{N}=2$ non-BPS, belonging to the orbit ($\mathds{R}%
\times \mathds{R}\equiv \mathds{R}^{2}$) \cite{ICL-2}
\begin{equation}
\mathcal{O}_{\mathcal{N}=2,STU,\mathbf{C1~}\left[ nBPS\right] }\sim \frac{%
\left[ \SL\left( 2,\mathds{R}\right) \right] ^{3}}{\mathds{R}^{2}}=\SL\left( 2,%
\mathds{R}\right) \times \left. \frac{\SO\left( 2,n\right) }{\SO\left(
n-1\right) \rtimes \left( \mathds{R}^{n-1}\times \mathds{R}\right) }\right|
_{n=2}.
\end{equation}
According to (\ref{ddd-1}), the supersymmetry reduction from maximal
$D=4$ supergravity reads
\begin{equation}
\mathcal{O}_{\mathcal{N}=8,\frac{1}{8}-BPS,\text{small}}\longrightarrow
\mathcal{O}_{\mathcal{N}=4,n_{V=6},\mathbf{C1~}\left[ \frac{1}{4}-BPS\right]
}\longrightarrow \mathcal{O}_{\mathcal{N}=2,STU,\mathbf{C1~}\left[ nBPS%
\right] }.
\end{equation}
The $\mathcal{N}=4$ origin is also confirmed by the symmetry enhancement due
to the fact that in solutions $\left[ \mathbf{1.i}\right] $-$\left[ \mathbf{%
1.iii}\right] $ two $\left| z_{i}\right| $'s are equal:
\begin{equation}
\left[ \SU\left( 2\right) \right] ^{4}\longrightarrow \left[ \SU\left(
2\right) \right] ^{2}\times \USp\left( 4\right) \sim \SO\left( 4\right) \times
\SO\left( 5\right) =mcs\left( \SO\left( 4,5\right) \right) .  \label{symm-en-1}
\end{equation}
Note that
\begin{equation}
\left| z_{a}\right| -\left| z_{1}\right| >0,~\forall a=1,2,3,  \label{rrel-1}
\end{equation}
all along $\mathcal{O}_{\mathcal{N}=2,STU,\mathbf{C1~}\left[
nBPS\right] }$. In terms of invariants of the $STU$ model, constraint
(\ref{rrel-1}) can be written as (see notation used in
\cite{Ceresole:2010nm}, and Refs. therein)
\begin{equation}
i_{\mathbf{a}}-i_{1}>0,~\forall \mathbf{a}=s,t,u.
\end{equation}
Solutions $\left[ \mathbf{1.i}\right] $-$\left[ \mathbf{1.iii}\right] $
correspond to the three solutions with maximal symmetry (in which two $%
\left| z_{i}\right| $'s are equal out of three) of (\ref{3-charge-sols})
with $i=1$, reading
\begin{equation}
\left\{
\begin{array}{l}
z_{1}=0; \\
\\
\multicolumn{1}{c}{\left| z_{2}\right| ^{4}+\left| z_{3}\right| ^{4}+\left|
z_{4}\right| ^{4}-2\left( \left| z_{2}\right| ^{2}\left| z_{3}\right|
^{2}+\left| z_{2}\right| ^{2}\left| z_{4}\right| ^{2}+\left| z_{3}\right|
^{2}\left| z_{4}\right| ^{2}\right) =0.}
\end{array}
\right.
\end{equation}
$\left[ \mathbf{1.i}\right] $-$\left[ \mathbf{1.iii}\right] $ are maximally
symmetric solutions respectively to the three ways in which the latter
quartic constraint can be rewritten:
\begin{align}
& \left| z_{2}\right| ^{4}+\left| z_{3}\right| ^{4}+\left| z_{4}\right|
^{4}-2\left( \left| z_{2}\right| ^{2}\left| z_{3}\right| ^{2}+\left|
z_{2}\right| ^{2}\left| z_{4}\right| ^{2}+\left| z_{3}\right| ^{2}\left|
z_{4}\right| ^{2}\right)  \notag \\
& =\left\{
\begin{array}{cc}
\left[ \mathbf{1.i}\right] : & \left( \left| z_{2}\right| ^{2}-\left|
z_{3}\right| ^{2}-\left| z_{4}\right| ^{2}\right) ^{2}=4\left| z_{3}\right|
^{2}\left| z_{4}\right| ^{2}; \\
\mathit{or} & ~ \\
\left[ \mathbf{1.ii}\right] : & \left( -\left| z_{2}\right| ^{2}+\left|
z_{3}\right| ^{2}-\left| z_{4}\right| ^{2}\right) ^{2}=4\left| z_{2}\right|
^{2}\left| z_{4}\right| ^{2}; \\
\mathit{or} & ~ \\
\left[ \mathbf{1.iii}\right] : & \left( -\left| z_{2}\right| ^{2}-\left|
z_{3}\right| ^{2}+\left| z_{4}\right| ^{2}\right) ^{2}=4\left| z_{2}\right|
^{2}\left| z_{3}\right| ^{2}.
\end{array}
\right.  \label{constr-1}
\end{align}

\item  The second prototypical representative solution reads
\begin{equation}
\left[ \mathbf{2.i}\right] :\left\{
\begin{array}{l}
z_{2}=0; \\
\\
2\left| z_{3}\right| =2\left| z_{4}\right| =\left| z_{1}\right| .
\end{array}
\right.
\end{equation}
This is a ($\frac{1}{2}$-)BPS solution: it is the $z_{2}=0$ limit of the
representative solution of orbit $\mathcal{O}_{\mathcal{N}=2,STU,\mathbf{C1~}%
\left[ BPS\right] }$, given by the ($\mathcal{N}=2$ $STU$ analogue
of the) of the solution treated at point C1) of Subsect. 4.4 of
\cite{Andrianopoli:2010bj}. By applying triality transformation
$\tau $, from $\left[ \mathbf{2.i}\right] $ one can generate other
two equivalent $\mathcal{N}=2$ BPS solutions, belonging to
$\mathcal{O}_{\mathcal{N}=2,STU,\mathbf{C1~}\left[ BPS\right] }$
itself, namely:
\begin{eqnarray}
\left[ \mathbf{2.ii}\right] &:&\left\{
\begin{array}{l}
z_{3}=0; \\
\\
2\left| z_{4}\right| =2\left| z_{2}\right| =\left| z_{1}\right| .
\end{array}
\right. \\
\left[ \mathbf{2.iii}\right] &:&\left\{
\begin{array}{l}
z_{4}=0; \\
\\
2\left| z_{2}\right| =2\left| z_{3}\right| =\left| z_{1}\right| .
\end{array}
\right.
\end{eqnarray}
Solutions $\left[ \mathbf{2.i}\right] $-$\left[ \mathbf{2.iii}\right] $ are $%
\tau $-equivalent, because it holds that:
\begin{equation}
\left[ \mathbf{2.i}\right] \overset{\tau }{\longrightarrow }\left[ \mathbf{%
2.ii}\right] \overset{\tau }{\longrightarrow }\left[ \mathbf{2.iii}\right]
\overset{\tau }{\longrightarrow }\left[ \mathbf{2.i}\right] .
\end{equation}
They all are $\mathcal{N}=2$ BPS, belonging to the orbit \cite{ICL-2}
\begin{equation}
\mathcal{O}_{\mathcal{N}=2,STU,\mathbf{C1~}\left[ BPS\right] }\sim \frac{%
\left[ \SL\left( 2,\mathds{R}\right) \right] ^{3}}{\mathds{R}^{2}}=\SL\left( 2,%
\mathds{R}\right) \times \left. \frac{\SO\left( 2,n\right) }{\SO\left(
n-1\right) \rtimes \left( \mathds{R}^{n-1}\times \mathds{R}\right) }\right|
_{n=2}.
\end{equation}
According to (\ref{ddd-1}), the supersymmetry reduction from maximal
$D=4$ supergravity reads
\begin{equation}
\mathcal{O}_{\mathcal{N}=8,\frac{1}{8}-BPS,\text{small}}\longrightarrow
\mathcal{O}_{\mathcal{N}=4,n_{V=6},\mathbf{C1~}\left[
\frac{1}{4}-BPS\right] }\longrightarrow
\mathcal{O}_{\mathcal{N}=2,STU,\mathbf{C1~}\left[ BPS\right] }.
\end{equation}
The $\mathcal{N}=4$ origin is also confirmed by the symmetry enhancement (%
\ref{symm-en-1}), due to the fact that in solutions $\left[ \mathbf{2.i}%
\right] $-$\left[ \mathbf{2.iii}\right] $ two $\left| z_{i}\right| $'s are
equal. Respectively, solutions $\left[ \mathbf{2.i}\right] $, $\left[
\mathbf{2.ii}\right] $ and $\left[ \mathbf{2.iii}\right] $ are solutions
with maximal symmetry (with $\left| z_{3}\right| =\left| z_{4}\right| $, $%
\left| z_{2}\right| =\left| z_{4}\right| $, and $\left| z_{2}\right| =\left|
z_{3}\right| $) of the following re-writings of (\ref{3-charge-sols}) for $%
i=2$, $i=3$ and $i=4$:
\begin{eqnarray}
\left[ \mathbf{2.i}\right] &:&\left\{
\begin{array}{l}
z_{2}=0; \\
\\
\multicolumn{1}{c}{
\begin{array}{c}
\left| z_{1}\right| ^{4}+\left| z_{3}\right| ^{4}+\left| z_{4}\right|
^{4}-2\left( \left| z_{1}\right| ^{2}\left| z_{3}\right| ^{2}+\left|
z_{1}\right| ^{2}\left| z_{4}\right| ^{2}+\left| z_{3}\right| ^{2}\left|
z_{4}\right| ^{2}\right) =0; \\
\Updownarrow \\
-\left| z_{1}\right| ^{2}+\left| z_{3}\right| ^{2}+\left| z_{4}\right|
^{2}=-2\left| z_{3}\right| \left| z_{4}\right| .
\end{array}
}
\end{array}
\right.  \notag \\
&& \\
\left[ \mathbf{2.ii}\right] &:&\left\{
\begin{array}{l}
z_{3}=0; \\
\\
\multicolumn{1}{c}{
\begin{array}{c}
\left| z_{1}\right| ^{4}+\left| z_{2}\right| ^{4}+\left| z_{4}\right|
^{4}-2\left( \left| z_{1}\right| ^{2}\left| z_{2}\right| ^{2}+\left|
z_{1}\right| ^{2}\left| z_{4}\right| ^{2}+\left| z_{2}\right| ^{2}\left|
z_{4}\right| ^{2}\right) =0; \\
\Updownarrow \\
-\left| z_{1}\right| ^{2}+\left| z_{2}\right| ^{2}+\left| z_{4}\right|
^{2}=-2\left| z_{2}\right| \left| z_{4}\right| .
\end{array}
}
\end{array}
\right.  \notag \\
&& \\
\left[ \mathbf{2.iii}\right] &:&\left\{
\begin{array}{l}
z_{4}=0; \\
\\
\multicolumn{1}{c}{
\begin{array}{c}
\left| z_{1}\right| ^{4}+\left| z_{2}\right| ^{4}+\left| z_{3}\right|
^{4}-2\left( \left| z_{1}\right| ^{2}\left| z_{2}\right| ^{2}+\left|
z_{1}\right| ^{2}\left| z_{3}\right| ^{2}+\left| z_{2}\right| ^{2}\left|
z_{3}\right| ^{2}\right) =0; \\
\Updownarrow \\
-\left| z_{1}\right| ^{2}+\left| z_{2}\right| ^{2}+\left| z_{3}\right|
^{2}=-2\left| z_{2}\right| \left| z_{3}\right| .
\end{array}
}
\end{array}
\right.  \notag \\
&&
\end{eqnarray}
Notice that the two sets $\left[ \mathbf{1.i}\right] $-$\left[ \mathbf{1.iii}%
\right] $ and $\left[ \mathbf{2.i}\right] $-$\left[ \mathbf{2.iii}\right] $
of $\tau $-equivalent solutions are related by the exchange $%
z_{1}\longleftrightarrow z_{2}$. This is irrelevant in $\mathcal{N}=4$
supersymmetry, because it amounts to exchanging the two skew-eigenvalues of
the $\mathcal{N}=4$ central charge matrix $Z_{AB}$. However, it matters in $%
\mathcal{N}=2$, because in this case it amounts to exchanging the $\mathcal{N%
}=2$ central charge with one of the three \textit{matter charges} (in $STU$
model, by triality). Both (isomorphic) orbits $\mathcal{O}_{\mathcal{N}%
=2,STU,\mathbf{C1~}\left[ \frac{1}{2}-BPS\right] }$ and $\mathcal{O}_{%
\mathcal{N}=2,STU,\mathbf{C1~}\left[ nBPS\right] }$ (see Eq. (\ref{ddd-2})-(%
\ref{ddd-3})) uplift to the $\mathcal{N}=4$ orbit $\mathcal{O}_{\mathcal{N}%
=4,n_{V=6},\mathbf{C1~}\left[ \frac{1}{4}-BPS\right] }$, as given by the
scheme (\ref{ddd-1}).

\item  The third prototypical representative solution reads
\begin{equation}
\left[ \mathbf{3.i}\right] :\left\{
\begin{array}{l}
z_{3}=0; \\
\\
2\left| z_{1}\right| =2\left| z_{2}\right| =\left| z_{4}\right| .
\end{array}
\right.
\end{equation}
This is a non-BPS solution: it is the representative solution of orbit $%
\mathcal{O}_{\mathcal{N}=2,STU,\mathbf{C2~}\left[ nBPS\right] }$,
namely the ($\mathcal{N}=2$ $STU$ analogue of the) representative
solution discussed at point C2) of Subsect. 4.4 of
\cite{Andrianopoli:2010bj}. In particular, $\left[
\mathbf{3.i}\right] $ corresponds to the branch
\begin{equation}
\left| z_{4}\right| -\left| z_{3}\right| =i_{u}-i_{t}>0
\end{equation}
of $\mathcal{O}_{\mathcal{N}=2,STU,\mathbf{C2~}\left[ nBPS\right] }$
itself (see \cite{Ceresole:2010nm}, also for notations of
invariants). By applying triality transformation $\tau $, from
$\left[ \mathbf{3.i}\right] $ one can generate
other two equivalent $\mathcal{N}=2$ non-BPS solutions, belonging to $%
\mathcal{O}_{\mathcal{N}=2,STU,\mathbf{C2~}\left[ nBPS\right] }$
itself, namely:
\begin{eqnarray}
\left[ \mathbf{3.ii}\right] &:&\left\{
\begin{array}{l}
z_{4}=0; \\
\\
2\left| z_{1}\right| =2\left| z_{3}\right| =\left| z_{2}\right| .
\end{array}
\right. \\
\left[ \mathbf{3.iii}\right] &:&\left\{
\begin{array}{l}
z_{2}=0; \\
\\
2\left| z_{1}\right| =2\left| z_{4}\right| =\left| z_{3}\right| .
\end{array}
\right.
\end{eqnarray}
Solutions $\left[ \mathbf{3.i}\right] $-$\left[ \mathbf{3.iii}\right] $ are $%
\tau $-equivalent, because it holds that:
\begin{equation}
\left[ \mathbf{3.i}\right] \overset{\tau }{\longrightarrow }\left[ \mathbf{%
3.ii}\right] \overset{\tau }{\longrightarrow }\left[ \mathbf{3.iii}\right]
\overset{\tau }{\longrightarrow }\left[ \mathbf{3.i}\right] .
\end{equation}
They all are $\mathcal{N}=2$ non-BPS, belonging to the orbit \cite{ICL-2}
\begin{equation}
\mathcal{O}_{\mathcal{N}=2,STU,\mathbf{C2~}\left[ nBPS\right] }\sim \frac{%
\left[ \SL\left( 2,\mathds{R}\right) \right] ^{3}}{\mathds{R}^{2}}=\SL\left( 2,%
\mathds{R}\right) \times \left. \frac{\SO\left( 2,n\right) }{\SO\left(
1,n-2\right) \rtimes \left( \mathds{R}^{1,n-2}\times \mathds{R}\right) }%
\right| _{n=2}.  \label{PA-1}
\end{equation}
According to (\ref{ddd-1}), the supersymmetry reduction from maximal
$D=4$ supergravity reads
\begin{equation}
\mathcal{O}_{\mathcal{N}=8,\frac{1}{8}-BPS,\text{small}}\longrightarrow
\mathcal{O}_{\mathcal{N}=4,n_{V=6},\mathbf{C2~}\left[ nBPS\right]
}\longrightarrow \mathcal{O}_{\mathcal{N}=2,STU,\mathbf{C2~}\left[ nBPS%
\right] }.  \label{SUSY-p-1}
\end{equation}
The $\mathcal{N}=4$ origin is also confirmed by the symmetry enhancement (%
\ref{symm-en-1}), due to the fact that in solutions $\left[ \mathbf{3.i}%
\right] $-$\left[ \mathbf{3.iii}\right] $ two $\left| z_{i}\right| $'s are
equal. Respectively, solutions $\left[ \mathbf{3.i}\right] $, $\left[
\mathbf{3.ii}\right] $ and $\left[ \mathbf{3.iii}\right] $ are solutions
with maximal symmetry (with $\left| z_{1}\right| =\left| z_{2}\right| $, $%
\left| z_{1}\right| =\left| z_{3}\right| $, and $\left| z_{1}\right| =\left|
z_{4}\right| $) of the following re-writings of (\ref{3-charge-sols}) for $%
i=3$, $i=4$ and $i=2$:
\begin{eqnarray}
\left[ \mathbf{3.i}\right] &:&\left\{
\begin{array}{l}
z_{3}=0; \\
\\
\multicolumn{1}{c}{
\begin{array}{c}
\left| z_{1}\right| ^{4}+\left| z_{2}\right| ^{4}+\left| z_{4}\right|
^{4}-2\left( \left| z_{1}\right| ^{2}\left| z_{2}\right| ^{2}+\left|
z_{1}\right| ^{2}\left| z_{4}\right| ^{2}+\left| z_{2}\right| ^{2}\left|
z_{4}\right| ^{2}\right) =0; \\
\Updownarrow \\
\left| z_{1}\right| ^{2}+\left| z_{2}\right| ^{2}-\left| z_{4}\right|
^{2}=-2\left| z_{1}\right| \left| z_{2}\right| .
\end{array}
}
\end{array}
\right.  \notag \\
&& \\
\left[ \mathbf{3.ii}\right] &:&\left\{
\begin{array}{l}
z_{4}=0; \\
\\
\multicolumn{1}{c}{
\begin{array}{c}
\left| z_{1}\right| ^{4}+\left| z_{2}\right| ^{4}+\left| z_{3}\right|
^{4}-2\left( \left| z_{1}\right| ^{2}\left| z_{2}\right| ^{2}+\left|
z_{1}\right| ^{2}\left| z_{3}\right| ^{2}+\left| z_{2}\right| ^{2}\left|
z_{3}\right| ^{2}\right) =0; \\
\Updownarrow \\
\left| z_{1}\right| ^{2}-\left| z_{2}\right| ^{2}+\left| z_{3}\right|
^{2}=-2\left| z_{1}\right| \left| z_{3}\right| .
\end{array}
}
\end{array}
\right.  \notag \\
&& \\
\left[ \mathbf{3.iii}\right] &:&\left\{
\begin{array}{l}
z_{2}=0; \\
\\
\multicolumn{1}{c}{
\begin{array}{c}
\left| z_{1}\right| ^{4}+\left| z_{3}\right| ^{4}+\left| z_{4}\right|
^{4}-2\left( \left| z_{1}\right| ^{2}\left| z_{3}\right| ^{2}+\left|
z_{1}\right| ^{2}\left| z_{4}\right| ^{2}+\left| z_{3}\right| ^{2}\left|
z_{4}\right| ^{2}\right) =0; \\
\Updownarrow \\
\left| z_{1}\right| ^{2}-\left| z_{3}\right| ^{2}+\left| z_{4}\right|
^{2}=-2\left| z_{1}\right| \left| z_{4}\right| .
\end{array}
}
\end{array}
\right.  \notag \\
&&
\end{eqnarray}

\item  The fourth prototypical representative solution reads
\begin{equation}
\left[ \mathbf{4.i}\right] :\left\{
\begin{array}{l}
z_{4}=0; \\
\\
2\left| z_{1}\right| =2\left| z_{2}\right| =\left| z_{3}\right| .
\end{array}
\right.
\end{equation}
This is another representative solution of orbit $\mathcal{O}_{\mathcal{N}%
=2,STU,\mathbf{C2~}\left[ nBPS\right] }$, namely of the ($\mathcal{N}=2$ $%
STU $ analogue of the) representative solution discussed at point
C2) of Subsect. 4.4 of \cite{Andrianopoli:2010bj}. In particular,
$\left[ \mathbf{4.i}\right] $ corresponds to the branch
\begin{equation}
\left| z_{3}\right| -\left| z_{4}\right| =i_{t}-i_{u}>0
\end{equation}
of $\mathcal{O}_{\mathcal{N}=2,STU,\mathbf{C2~}\left[ nBPS\right] }$
itself (see \cite{Ceresole:2010nm}, also for notations of
invariants). By applying triality transformation $\tau $, from
$\left[ \mathbf{4.i}\right] $ one can generate
other two equivalent $\mathcal{N}=2$ non-BPS solutions, belonging to $%
\mathcal{O}_{\mathcal{N}=2,STU,\mathbf{C2~}\left[ nBPS\right] }$
itself, namely:
\begin{eqnarray}
\left[ \mathbf{4.ii}\right] &:&\left\{
\begin{array}{l}
z_{2}=0; \\
\\
2\left| z_{1}\right| =2\left| z_{3}\right| =\left| z_{4}\right| .
\end{array}
\right. \\
\left[ \mathbf{4.iii}\right] &:&\left\{
\begin{array}{l}
z_{3}=0; \\
\\
2\left| z_{1}\right| =2\left| z_{4}\right| =\left| z_{2}\right| .
\end{array}
\right.
\end{eqnarray}
Solutions $\left[ \mathbf{4.i}\right] $-$\left[ \mathbf{4.iii}\right] $ are $%
\tau $-equivalent, because it holds that:
\begin{equation}
\left[ \mathbf{4.i}\right] \overset{\tau }{\longrightarrow }\left[ \mathbf{%
4.ii}\right] \overset{\tau }{\longrightarrow }\left[ \mathbf{4.iii}\right]
\overset{\tau }{\longrightarrow }\left[ \mathbf{4.i}\right] .
\end{equation}
They all are $\mathcal{N}=2$ non-BPS, belonging to the orbit
(\ref{PA-1}), with supersymmetry reduction from maximal $D=4$
supergravity given by (\ref{SUSY-p-1}). The $\mathcal{N}=4$ origin
is also confirmed by the symmetry enhancement (\ref{symm-en-1}), due
to the fact that in solutions $\left[ \mathbf{4.i}\right] $-$\left[
\mathbf{4.iii}\right] $ two $\left|
z_{i}\right| $'s are equal. Respectively, solutions $\left[ \mathbf{4.i}%
\right] $, $\left[ \mathbf{4.ii}\right] $ and $\left[ \mathbf{4.iii}\right] $
are solutions with maximal symmetry (with $\left| z_{1}\right| =\left|
z_{2}\right| $, $\left| z_{1}\right| =\left| z_{3}\right| $, and $\left|
z_{1}\right| =\left| z_{4}\right| $) of the following re-writings of (\ref
{3-charge-sols}) for $i=4$, $i=2$ and $i=3$:
\begin{eqnarray}
\left[ \mathbf{4.i}\right] &:&\left\{
\begin{array}{l}
z_{4}=0; \\
\\
\multicolumn{1}{c}{
\begin{array}{c}
\left| z_{1}\right| ^{4}+\left| z_{2}\right| ^{4}+\left| z_{3}\right|
^{4}-2\left( \left| z_{1}\right| ^{2}\left| z_{2}\right| ^{2}+\left|
z_{1}\right| ^{2}\left| z_{3}\right| ^{2}+\left| z_{2}\right| ^{2}\left|
z_{3}\right| ^{2}\right) =0; \\
\Updownarrow \\
\left| z_{1}\right| ^{2}+\left| z_{2}\right| ^{2}-\left| z_{3}\right|
^{2}=-2\left| z_{1}\right| \left| z_{2}\right| .
\end{array}
}
\end{array}
\right.  \notag \\
&& \\
\left[ \mathbf{4.ii}\right] &:&\left\{
\begin{array}{l}
z_{2}=0; \\
\\
\multicolumn{1}{c}{
\begin{array}{c}
\left| z_{1}\right| ^{4}+\left| z_{3}\right| ^{4}+\left| z_{4}\right|
^{4}-2\left( \left| z_{1}\right| ^{2}\left| z_{3}\right| ^{2}+\left|
z_{1}\right| ^{2}\left| z_{4}\right| ^{2}+\left| z_{3}\right| ^{2}\left|
z_{4}\right| ^{2}\right) =0; \\
\Updownarrow \\
\left| z_{1}\right| ^{2}+\left| z_{3}\right| ^{2}-\left| z_{4}\right|
^{2}=-2\left| z_{1}\right| \left| z_{3}\right| .
\end{array}
}
\end{array}
\right.  \notag \\
&& \\
\left[ \mathbf{4.iii}\right] &:&\left\{
\begin{array}{l}
z_{3}=0; \\
\\
\multicolumn{1}{c}{
\begin{array}{c}
\left| z_{1}\right| ^{4}+\left| z_{2}\right| ^{4}+\left| z_{4}\right|
^{4}-2\left( \left| z_{1}\right| ^{2}\left| z_{2}\right| ^{2}+\left|
z_{1}\right| ^{2}\left| z_{4}\right| ^{2}+\left| z_{2}\right| ^{2}\left|
z_{4}\right| ^{2}\right) =0; \\
\Updownarrow \\
\left| z_{1}\right| ^{2}-\left| z_{2}\right| ^{2}+\left| z_{4}\right|
^{2}=-2\left| z_{1}\right| \left| z_{4}\right| .
\end{array}
}
\end{array}
\right.  \notag \\
&&
\end{eqnarray}
Notice that the two sets $\left[ \mathbf{3.i}\right] $-$\left[ \mathbf{3.iii}%
\right] $ and $\left[ \mathbf{4.i}\right] $-$\left[ \mathbf{4.iii}\right] $
of $\tau $-equivalent solutions are related by the exchange $%
z_{3}\longleftrightarrow z_{4}$, immaterial both in $\mathcal{N}=4$
supergravity (due to $\mathcal{N}=4$ supersymmetry) and in
$\mathcal{N}=2$ $STU$ model (for \textit{triality}
symmetry).\bigskip
\end{enumerate}

By noting that the four independent solutions given (\ref{3-charge-sols}) (%
\textit{e.g.} for $i=1,2,3,4$) are related though the iterated action of
\textit{quaterniality} permutation symmetry $\pi $ defined in (\ref{pi-def}%
), and by recalling definitions (\ref{tau-def}) and (\ref{pi-def}), one can
determine how the twelve representative solutions of type $\mathbf{1}$, $%
\mathbf{2}$, $\mathbf{3}$ and $\mathbf{4}$ treated above are related through
(composition of) $\tau $ and $\pi $. One can present the resulting web of
relations in four equivalent ways, corresponding to using each of the four
sets $\mathbf{1}$, $\mathbf{2}$, $\mathbf{3}$ and $\mathbf{4}$ of three $%
\tau $-equivalent representative solutions as the ``pivot'' (first column on the
left) for the iterated application of (composition(s) of) $\tau $ and $\pi $%
:
\begin{gather}
\mathbf{I}\equiv \text{set~}\mathbf{1~}\text{as~\textit{``pivot''}}:  \notag
\\
\fbox{$
\begin{array}{ccccccccc}
\left[ \mathbf{1.i}\right] & \overset{\pi }{\longrightarrow } & \left[
\mathbf{3.iii}\right] & \overset{\pi }{\longrightarrow } & \left[ \mathbf{3.i%
}\right] & \overset{\pi }{\longrightarrow } & \left[ \mathbf{2.iii}\right] &
\overset{\pi }{\longrightarrow } & \left[ \mathbf{1.i}\right] \\
\downarrow \tau &  &  &  &  &  &  &  & \downarrow \tau \\
\left[ \mathbf{1.ii}\right] & \overset{\pi }{\longrightarrow } & \left[
\mathbf{4.ii}\right] & \overset{\pi }{\longrightarrow } & \left[ \mathbf{2.ii%
}\right] & \overset{\pi }{\longrightarrow } & \left[ \mathbf{3.ii}\right] &
\overset{\pi }{\longrightarrow } & \left[ \mathbf{1.ii}\right] \\
\downarrow \tau &  &  &  &  &  &  &  & \downarrow \tau \\
\left[ \mathbf{1.iii}\right] & \overset{\pi }{\longrightarrow } & \left[
\mathbf{2.i}\right] & \overset{\pi }{\longrightarrow } & \left[ \mathbf{4.iii%
}\right] & \overset{\pi }{\longrightarrow } & \left[ \mathbf{4.i}\right] &
\overset{\pi }{\longrightarrow } & \left[ \mathbf{1.iii}\right] \\
\downarrow \tau &  &  &  &  &  &  &  & \downarrow \tau \\
\left[ \mathbf{1.i}\right] & \overset{\pi }{\longrightarrow } & \left[
\mathbf{3.iii}\right] & \overset{\pi }{\longrightarrow } & \left[ \mathbf{3.i%
}\right] & \overset{\pi }{\longrightarrow } & \left[ \mathbf{2.iii}\right] &
\overset{\pi }{\longrightarrow } & \left[ \mathbf{1.i}\right]
\end{array}
$}
\end{gather}
\begin{gather}
\mathbf{II}\equiv \text{set~}\mathbf{2~}\text{as~\textit{``pivot''}}:  \notag
\\
\fbox{$
\begin{array}{ccccccccc}
\left[ \mathbf{2.i}\right] & \overset{\pi }{\longrightarrow } & \left[
\mathbf{4.iii}\right] & \overset{\pi }{\longrightarrow } & \left[ \mathbf{4.i%
}\right] & \overset{\pi }{\longrightarrow } & \left[ \mathbf{1.iii}\right] &
\overset{\pi }{\longrightarrow } & \left[ \mathbf{2.i}\right] \\
\downarrow \tau &  &  &  &  &  &  &  & \downarrow \tau \\
\left[ \mathbf{2.ii}\right] & \overset{\pi }{\longrightarrow } & \left[
\mathbf{3.ii}\right] & \overset{\pi }{\longrightarrow } & \left[ \mathbf{1.ii%
}\right] & \overset{\pi }{\longrightarrow } & \left[ \mathbf{4.ii}\right] &
\overset{\pi }{\longrightarrow } & \left[ \mathbf{2.ii}\right] \\
\downarrow \tau &  &  &  &  &  &  &  & \downarrow \tau \\
\left[ \mathbf{2.iii}\right] & \overset{\pi }{\longrightarrow } & \left[
\mathbf{1.i}\right] & \overset{\pi }{\longrightarrow } & \left[ \mathbf{3.iii%
}\right] & \overset{\pi }{\longrightarrow } & \left[ \mathbf{3.i}\right] &
\overset{\pi }{\longrightarrow } & \left[ \mathbf{2.iii}\right] \\
\downarrow \tau &  &  &  &  &  &  &  & \downarrow \tau \\
\left[ \mathbf{2.i}\right] & \overset{\pi }{\longrightarrow } & \left[
\mathbf{4.iii}\right] & \overset{\pi }{\longrightarrow } & \left[ \mathbf{4.i%
}\right] & \overset{\pi }{\longrightarrow } & \left[ \mathbf{1.iii}\right] &
\overset{\pi }{\longrightarrow } & \left[ \mathbf{2.i}\right]
\end{array}
$}
\end{gather}
\begin{gather}
\mathbf{III}\equiv \text{set~}\mathbf{3~}\text{as~\textit{``pivot''}}:
\notag \\
\fbox{$
\begin{array}{ccccccccc}
\left[ \mathbf{3.i}\right] & \overset{\pi }{\longrightarrow } & \left[
\mathbf{2.iii}\right] & \overset{\pi }{\longrightarrow } & \left[ \mathbf{1.i%
}\right] & \overset{\pi }{\longrightarrow } & \left[ \mathbf{3.iii}\right] &
\overset{\pi }{\longrightarrow } & \left[ \mathbf{3.i}\right] \\
\downarrow \tau &  &  &  &  &  &  &  & \downarrow \tau \\
\left[ \mathbf{3.ii}\right] & \overset{\pi }{\longrightarrow } & \left[
\mathbf{1.ii}\right] & \overset{\pi }{\longrightarrow } & \left[ \mathbf{4.ii%
}\right] & \overset{\pi }{\longrightarrow } & \left[ \mathbf{2.ii}\right] &
\overset{\pi }{\longrightarrow } & \left[ \mathbf{3.ii}\right] \\
\downarrow \tau &  &  &  &  &  &  &  & \downarrow \tau \\
\left[ \mathbf{3.iii}\right] & \overset{\pi }{\longrightarrow } & \left[
\mathbf{3.i}\right] & \overset{\pi }{\longrightarrow } & \left[ \mathbf{2.iii%
}\right] & \overset{\pi }{\longrightarrow } & \left[ \mathbf{1.i}\right] &
\overset{\pi }{\longrightarrow } & \left[ \mathbf{3.iii}\right] \\
\downarrow \tau &  &  &  &  &  &  &  & \downarrow \tau \\
\left[ \mathbf{3.i}\right] & \overset{\pi }{\longrightarrow } & \left[
\mathbf{2.iii}\right] & \overset{\pi }{\longrightarrow } & \left[ \mathbf{1.i%
}\right] & \overset{\pi }{\longrightarrow } & \left[ \mathbf{3.iii}\right] &
\overset{\pi }{\longrightarrow } & \left[ \mathbf{3.i}\right]
\end{array}
$}
\end{gather}
\begin{gather}
\mathbf{IV}\equiv \text{set~}\mathbf{4~}\text{as~\textit{``pivot''}}:  \notag
\\
\fbox{$
\begin{array}{ccccccccc}
\left[ \mathbf{4.i}\right] & \overset{\pi }{\longrightarrow } & \left[
\mathbf{1.iii}\right] & \overset{\pi }{\longrightarrow } & \left[ \mathbf{2.i%
}\right] & \overset{\pi }{\longrightarrow } & \left[ \mathbf{4.iii}\right] &
\overset{\pi }{\longrightarrow } & \left[ \mathbf{4.i}\right] \\
\downarrow \tau &  &  &  &  &  &  &  & \downarrow \tau \\
\left[ \mathbf{4.ii}\right] & \overset{\pi }{\longrightarrow } & \left[
\mathbf{2.ii}\right] & \overset{\pi }{\longrightarrow } & \left[ \mathbf{3.ii%
}\right] & \overset{\pi }{\longrightarrow } & \left[ \mathbf{1.ii}\right] &
\overset{\pi }{\longrightarrow } & \left[ \mathbf{4.ii}\right] \\
\downarrow \tau &  &  &  &  &  &  &  & \downarrow \tau \\
\left[ \mathbf{4.iii}\right] & \overset{\pi }{\longrightarrow } & \left[
\mathbf{4.i}\right] & \overset{\pi }{\longrightarrow } & \left[ \mathbf{1.iii%
}\right] & \overset{\pi }{\longrightarrow } & \left[ \mathbf{2.i}\right] &
\overset{\pi }{\longrightarrow } & \left[ \mathbf{4.iii}\right] \\
\downarrow \tau &  &  &  &  &  &  &  & \downarrow \tau \\
\left[ \mathbf{4.i}\right] & \overset{\pi }{\longrightarrow } & \left[
\mathbf{1.iii}\right] & \overset{\pi }{\longrightarrow } & \left[ \mathbf{2.i%
}\right] & \overset{\pi }{\longrightarrow } & \left[ \mathbf{4.iii}\right] &
\overset{\pi }{\longrightarrow } & \left[ \mathbf{4.i}\right]
\end{array}
$}
\end{gather}
Notice that the first and the last rows and the first and the last columns
of each array coincide, due to the idempotency properties of $\tau $ and $%
\pi $ themselves. Moreover, the rows of different arrays are related by
cyclical reshufflings. The analogous $\pi $-patterns of the arrays $\mathbf{I%
}$ and $\mathbf{II}$, as well as of arrays $\mathbf{III}$ and $\mathbf{IV}$,
can be explained through the different roles of $z_{1}$ within the
corresponding sets ($\mathbf{1}$ and $\mathbf{2}$, as well as $\mathbf{3}$
and $\mathbf{4}$, respectively) of $\tau $-equivalent representative
solutions (see treatment above).\textbf{\ }

In particular, the second row of each array has the remarkable property of
containing only representative solutions of the type $\left[ \mathbf{A.ii}%
\right] $ ($\mathbf{A}=\mathbf{1}$, $\mathbf{2}$, $\mathbf{3}$, $\mathbf{4}$%
), related through iterated application of the \textit{quaterniality}
permutation $\pi $. Thus, the four solutions of type $\left[ \mathbf{A.ii}%
\right] $ (with $\mathbf{A}=\mathbf{1}$, $\mathbf{2}$, $\mathbf{3}$ and $%
\mathbf{4}$, one ($\frac{1}{2}$-)BPS and three non-BPS in $\mathcal{N}=2$
supersymmetry) can conveniently and consistently be taken in one-to-one
correspondence with the four Young tableaux of the four $\SO_{0}\left(
4,4\right) $-nilpotent orbits of real dimension $16$, in turn related to the
four $A$-$W$ classes belonging to the family $L_{a_{2}0_{3}\oplus \overline{1}%
} $.

\subsubsection{$A$-$B$-EPR Classes $\Leftrightarrow L_{a_{2}b_{2}}$ : \textit{%
Critical} $\frac{1}{2}$-BPS and non-BPS Orbits}

The $2$-charge (\textit{critical}) ``small'' orbit of the $\left( \mathbf{2},%
\mathbf{2},\mathbf{2}\right) $ of $G_{4,STU}$ is given by (see Table
VI of \cite{Borsten:2009yb})
\begin{equation}
\mathcal{O}_{STU,crit.}=\frac{\left[ \SL\left( 2,\mathds{R}\right)
\right] ^{3}}{\SO\left( 2,1\right) \times
\mathds{R}},~\text{dim}_{\mathds{R}}=5. \label{stu-2-charge}
\end{equation}
$\mathcal{O}_{STU,crit.}$ is defined by an $\left[ \SL\left( 2,\mathds{R}%
\right) \right] ^{3}$-invariant set of constraints\ which involve
first-order functional derivatives of $\mathcal{I}_{4}$ itself:
\begin{equation}
\frac{\partial \mathcal{I}_{4}}{\partial z_{i}}=2z_{i}\overline{z_{i}}^{2}-2%
\overline{z_{i}}\sum_{j\neq i}\left| z_{j}\right| ^{2}+4\prod_{j\neq
i}z_{j}=0,~\forall i.  \label{2-charge-constraint}
\end{equation}
Note that, for each fixed $i$, $\frac{\partial \mathcal{I}_{4}}{\partial
z_{i}}$ is manifestly invariant under cyclic permutations of the index $%
j\neq i$. As a consequence, the whole set of four constraints (\ref
{2-charge-constraint}) is manifestly $\pi $-invariant.

The four constraints (\ref{2-charge-constraint}) admit six representative
solutions, namely:
\begin{equation}
\left\{
\begin{array}{ll}
\mathbf{I}: &
\begin{array}{l}
z_{1}=0=z_{2},~z_{3}=z_{4}\neq 0;
\end{array}
\\
\mathbf{II}: &
\begin{array}{l}
z_{1}=0=z_{3},~z_{2}=z_{4}\neq 0;
\end{array}
\\
\mathbf{III}: &
\begin{array}{l}
z_{1}=0=z_{4},~z_{2}=z_{3}\neq 0;
\end{array}
\\
\mathbf{IV}: &
\begin{array}{l}
z_{2}=0=z_{3},~z_{1}=z_{4}\neq 0;
\end{array}
\\
\mathbf{V}: &
\begin{array}{l}
z_{2}=0=z_{4},~z_{1}=z_{3}\neq 0;
\end{array}
\\
\mathbf{VI}: &
\begin{array}{l}
z_{3}=0=z_{4},~z_{1}=z_{2}\neq 0,
\end{array}
\end{array}
\right.  \label{2-charge-sols-mcs-expl}
\end{equation}
which can be split into two sets (each being separately $\pi $-invariant)
\begin{equation}
\begin{array}{cc}
\mathbf{I}\overset{\pi }{\rightarrow }\mathbf{IV\overset{\pi }{\rightarrow }%
VI\overset{\pi }{\rightarrow }III\overset{\pi }{\rightarrow }I}: & \left\{
\begin{array}{l}
z_{i}=z_{i+1}=0; \\
\\
z_{i+2}=z_{i+3}\neq 0;
\end{array}
\right. \\
~ & ~ \\
\mathbf{II\overset{\pi }{\rightarrow }V\overset{\pi }{\rightarrow }II}: &
\left\{
\begin{array}{l}
z_{i}=z_{i+2}=0; \\
\\
z_{i+1}=z_{i+3}\neq 0.
\end{array}
\right.
\end{array}
\label{2-charge-sols-mcs}
\end{equation}
The non-vanishing $z_{i}$'s given by solutions (\ref{2-charge-sols-mcs-expl}%
) are generally complex. The set (\ref{2-charge-sols-mcs-expl}) (or
equivalently (\ref{2-charge-sols-mcs})) exhibits the maximal compact
symmetry consistent with \cite{Ferrara:1997uz,Lu:1997bg}
\begin{equation}
\mathcal{O}_{\mathcal{N}=8,\frac{1}{4}-BPS}=\frac{E_{7\left( 7\right) }}{%
\left( \SO\left( 6,5\right) \rtimes \mathds{R}^{32}\right) \times \mathds{R}},
\end{equation}
namely $\SO\left( 6\right) \times \SO(5)\sim \SU(4)\times \USp\left( 4\right)
=mcs\left( \SO\left( 6,5\right) \right) $.

The properties under the action of $\pi $ indicated in (\ref
{2-charge-sols-mcs}) determine  two sets of Young tableaux, with
cardinality $4$ and $2$ respectively \cite{Borsten:2010db}. In each
of these two sets, the Young tableaux are related through $D=3$
permutation symmetry, and thus they can be identified up to $D=3$
permutations. The six representative solutions
(\ref{2-charge-sols-mcs}) are related to the six $A$-$B$-EPR classes
of $4$-qubits entanglement, organized in the two sets of Young
tableaux corresponding to two groupings $\SO_{0}\left( 4,4\right)
$-nilpotent orbits
of dimension $12$ (identified up to $\pi $), both associated to the family $%
L_{a_{2}b_{2}}$ of \cite{Verstraete:2002}.

Out of the six Young tableaux associated to $L_{a_{2}b_{2}}$, three
correspond to $\frac{1}{2}$-BPS $2$-charge $STU$ BHs, and three
correspond to non-BPS $2$-charges $STU$ BHs, with an equal amount of
supersymmetric and non-supersymmetric solutions in the two subsets
given in (\ref {2-charge-sols-mcs}). Such a $D=4$ supersymmetry
interpretation can be summarized by the following scheme
\cite{ICL-2}
\begin{equation}
\begin{array}{cccccc}
\underline{\mathcal{N}=8}: &  &  & \mathcal{O}_{\mathcal{N}=8,\frac{1}{4}%
-BPS} &  &  \\
&  & \swarrow & \downarrow & \searrow &  \\
\underline{\mathcal{N}=4}: & \mathcal{O}_{\mathcal{N}=4,n_{V=6},\mathbf{A1~}%
\left[ \frac{1}{2}-BPS\right] } &  & \mathcal{O}_{\mathcal{N}=4,n_{V=6},%
\mathbf{B~}\left[ \frac{1}{4}-BPS\right] } &  & \mathcal{O}_{\mathcal{N}%
=4,n_{V=6},\mathbf{A2~}\left[ nBPS\right] } \\
& \SL\left( 2,\mathds{R}\right) \times \frac{\SO\left( 6,6\right) }{\SO\left(
5,6\right) \times \mathds{R}} &  & \SL\left( 2,\mathds{R}\right) \times \frac{%
\SO\left( 6,6\right) }{\SO\left( 2,1\right) \times \SO\left( 4,4\right) \times
\mathds{R}} &  & \SL\left( 2,\mathds{R}\right) \times \frac{\SO\left(
6,6\right) }{\SO\left( 6,5\right) \times \mathds{R}} \\
& \downarrow &  & \downarrow &  & \downarrow \\
\underline{\mathcal{N}=2}: & \overset{\mathbf{VI:}}{\mathcal{O}_{\mathcal{N}%
=2,STU,\mathbf{A1~}\left[ \frac{1}{2}-BPS\right] }} &  &
\begin{array}{c}
\overset{\mathbf{IV},\mathbf{V:}}{\mathcal{O}_{\mathcal{N}=2,STU,\mathbf{B~}%
\left[ \frac{1}{2}-BPS\right] }} \\
\updownarrow \ast \\
\overset{\mathbf{II},\mathbf{III:}}{\mathcal{O}_{\mathcal{N}=2,STU,\mathbf{B~%
}\left[ nBPS\right] }}
\end{array}
&  &
\overset{\mathbf{I:}}{\mathcal{O}_{\mathcal{N}=2,STU,\mathbf{A2~}\left[
nBPS\right] }},
\end{array}
\label{dddd-1}
\end{equation}
where ``$\mathbf{A1}$'', ``$\mathbf{A2}$'' and ``$\mathbf{B}$'' refer to the
classification of \cite{Cerchiai:2009pi,Andrianopoli:2010bj}, and the Latin uppercase numbers
denote the solutions given in (\ref{2-charge-sols-mcs}). This is consistent
with the results of \cite{Cerchiai:2009pi} and \cite{Andrianopoli:2010bj}. It holds that
\begin{eqnarray}
\mathcal{O}_{\mathcal{N}=2,STU,\mathbf{A1~}\left[
\frac{1}{2}-BPS\right] }
&\sim &\mathcal{O}_{\mathcal{N}=2,STU,\mathbf{B~}\left[ \frac{1}{2}-BPS%
\right] }\sim \mathcal{O}_{\mathcal{N}=2,STU,\mathbf{B~}\left[
nBPS\right] }
\label{dddd-2} \\
&\sim &\mathcal{O}_{\mathcal{N}=2,STU,\mathbf{A2~}\left[ nBPS\right]
}\sim \mathcal{O}_{STU,crit.}=\frac{\left[ \SL\left(
2,\mathds{R}\right) \right]
^{3}}{\SO\left( 2,1\right) \times \mathds{R}},  \notag \\
&&  \label{dddd-3}
\end{eqnarray}
consistent with Table VI of \cite{Borsten:2009yb}. Consequently, Eqs. (\ref
{dddd-1})-(\ref{dddd-3}) correspond to the following chains of maximal
symmetric embeddings: for the numerators of cosets it holds (\ref{ch-1}),
whereas for the stabilizers it holds:
\begin{equation}
\left( \SO\left( 6,5\right) \rtimes \mathds{R}^{32}\right) \times \mathds{R}%
\supsetneq \left( \SO\left( 2,1\right) \times \SO\left( 4,4\right) \rtimes
\left( \mathds{R}^{\left( 2,8_{s}\right) }\times \mathds{R}^{\left(
2,8_{c}\right) }\right) \right) \times \mathds{R},  \label{ch-3}
\end{equation}
where the double-spinors $\mathds{R}^{\left( 2,8_{s}\right) }$ and $\mathds{R%
}^{\left( 2,8_{c}\right) }$ are truncated out.

Note that the sets $\left\{
\mathbf{IV},\mathbf{V},\mathbf{VI}\right\} $ and $\left\{
\mathbf{I},\mathbf{II},\mathbf{III}\right\} $ are separately
invariant under $\tau $. Since $\tau $ always commutes with $D=4$
supersymmetry, each set is characterised by a unique supersymmetry
property,
namely $\left\{ \mathbf{IV},\mathbf{V},\mathbf{VI}\right\} $ is $\frac{1}{2}$%
-BPS, whereas $\left\{ \mathbf{I},\mathbf{II},\mathbf{III}\right\} $ is
non-BPS. Thus, each of the two sets of Young tableaux corresponding to the
solutions (\ref{2-charge-sols-mcs}) has $50\%$ supersymmetric and $50\%$
non-supersymmetric contents; namely, the set $\left\{ \mathbf{I},\mathbf{III}%
,\mathbf{IV},\mathbf{VI}\right\} $ contains two $\frac{1}{2}$-BPS ($\mathbf{%
IV}$ and $\mathbf{VI}$) and two non-BPS ($\mathbf{I}$ and $\mathbf{III}$) $2$%
-charge solutions, whereas the set $\left\{ \mathbf{II},\mathbf{V}\right\} $
contains one $\frac{1}{2}$-BPS ($\mathbf{V}$) and one non-BPS ($\mathbf{II}$%
) $2$-charge solution.

One can also check that the $mcs$ of the non-translational part of
the stabilizer of $\mathcal{O}_{STU,crit.}$ is (non-maximally)
embedded into the
$mcs$'s of the non-translational part of the stabilizers of the two $%
\SO_{0}\left( 4,4\right) $-nilpotent orbits of dimension $12$, namely:
\begin{equation}
mcs\left( (S)O\left( 2,1\right) \right) =\SO\left( 2\right) \subsetneq
\left\{
\begin{array}{l}
\SO\left( 3\right) \times \SO\left( 2\right) =mcs\left( \SO\left( 3,2;\mathds{R}%
\right) \right) ; \\
\\
\SO\left( 4\right) =mcs\left( Sp\left( 4,\mathds{R}\right) \right) .
\end{array}
\right.
\end{equation}

\subsubsection{$A$-$B$-$C$-$D$ Class $\Leftrightarrow L_{abc_{2}}$ : \textit{%
Doubly-Critical} $\frac{1}{2}$-BPS Orbit}

The $1$-charge (\textit{doubly-critical}) ``small'' orbit of the
$\left( \mathbf{2},\mathbf{2},\mathbf{2}\right) $ of $G_{4,STU}$ is
given by (see Table VI of \cite{Borsten:2009yb})
\begin{equation}
\mathcal{O}_{STU,doubly-crit}=\frac{\left[ \SL\left( 2,\mathds{R}\right) %
\right] ^{3}}{\left[ \SO\left( 1,1\right) \right] ^{2}\rtimes \mathds{R}^{3}}%
,~\text{dim}_{\mathds{R}}=4.  \label{stu-1-charge}
\end{equation}
$\mathcal{O}_{STU,doubly-crit}$ is defined by an $\left[ \SL\left( 2,\mathds{R%
}\right) \right] ^{3}$-invariant set of constraints which involve suitable
projections of second-order functional derivatives of $\mathcal{I}_{4}$
itself (see \textit{e.g.} \cite{Cerchiai:2009pi,Ceresole:2010nm}). Such a set of constraints can
be recast in the following form:
\begin{equation}
\left\{
\begin{array}{l}
\left| z_{1}\right| ^{2}=\left| z_{2}\right| ^{2}=\left| z_{3}\right|
^{2}=\left| z_{4}\right| ^{2}; \\
\\
z_{i}z_{j}-\overline{z_{k}}\overline{z_{l}}=0,~\forall i\neq j\neq k\neq l.
\end{array}
\right.  \label{1-charge-constraint}
\end{equation}
Note the similarity of the second of (\ref{1-charge-constraint}) with the
Attractor Eqs. (\ref{N=8-AEs-gen}) themselves.

The constraints (\ref{1-charge-constraint}) are manifestly $\pi $-invariant.
It should also be pointed out that out of the second set of constraints of (%
\ref{1-charge-constraint}), only the following Eqs. are independent:
\begin{equation}
\left\{
\begin{array}{l}
z_{1}z_{2}-\overline{z_{3}}\overline{z_{4}}=0; \\
\\
z_{1}z_{3}-\overline{z_{2}}\overline{z_{4}}=0,
\end{array}
\right.
\end{equation}
and all the others can be obtained through iterated action of $\pi $ (and
through complex conjugation).

Constraints (\ref{1-charge-constraint}) admit the following representative
solutions, manifestly $\pi $-invariant:
\begin{equation}
\left\{
\begin{array}{l}
\left| z_{i}\right| =\eta e^{i\frac{\varphi }{4}},~\eta \in \mathds{R}%
_{0}^{+},~\forall i, \\
\\
\varphi =2m\pi ,~m\in \mathds{Z}.
\end{array}
\right.  \label{1-charge-sol}
\end{equation}
Notice the similarity of ``small'' $1$-charge (\textit{doubly-critical}) $%
\frac{1}{2}$-BPS (both in $\mathcal{N}=2$ and $\mathcal{N}=8$) solution (\ref
{1-charge-sol}) with the ``large'' (and thus attractor) (non-BPS $Z_{H}\neq
0 $ in $\mathcal{N}=2$ and non-BPS in $\mathcal{N}=8$) solution (\ref{nBPS}).

The solution (\ref{1-charge-sol}) exhibits the maximal compact symmetry
consistent with \cite{Ferrara:1997uz,Lu:1997bg}
\begin{equation}
\mathcal{O}_{\mathcal{N}=8,\frac{1}{2}-BPS}=\frac{E_{7\left( 7\right) }}{%
E_{6\left( 6\right) }\rtimes \mathds{R}^{27}},
\end{equation}
namely $\USp(8)=mcs\left( E_{6\left( 6\right) }\right) $.

The manifestly $\pi $-invariant solution (\ref{1-charge-sol})
corresponds to a unique Young tableaux, manifestly invariant under
$D=3$ permutation symmetry, and related to the totally separable
$A$-$B$-$C$-$D$ class of $4$-qubits
entanglement. This in turn determines a unique $\SO_{0}\left( 4,4\right) $%
-nilpotent orbit, namely the \textit{minimal} one of real dimension $10$,
corresponding to the family $L_{abc_{2}}$ of $4$-qubits entanglement states
\cite{Borsten:2010db}.

It generally holds that in $D=4$ the $1$-charge orbit is always
unique and maximally supersymmetric (namely, $\frac{1}{2}$-BPS): it
corresponds to the \textit{minimal} nilpotent $G_{3}$-orbit.

Correspondingly, there exists a unique ``small'' $1$-charge (\textit{%
doubly-critical}) $\frac{1}{2}$-BPS orbit in $\mathcal{N}=2$, $D=4$
$STU$ model, given by (\ref{stu-1-charge}). Such a $D=4$
supersymmetry
interpretation can be summarized as follows \cite{ICL-2} (subscript denote $%
\SO\left( 1,1\right) $-weights):
\begin{equation}
\begin{array}{cc}
\underline{\mathcal{N}=8}: & \mathcal{O}_{\mathcal{N}=8,\frac{1}{2}-BPS} \\
& \downarrow \\
\underline{\mathcal{N}=4}: &
\begin{array}{c}
\mathcal{O}_{\mathcal{N}=4,n_{V=6},\mathbf{A3}\left[ \frac{1}{2}-BPS\right] }
\\
\\
\SL\left( 2,\mathds{R}\right) \times \frac{\SO\left( 6,6\right) }{\left[
\SO\left( 1,1\right) \times \SO\left( 5,5\right) \right] \rtimes \left(
\mathds{R}^{5,5_{-2}}\times \mathds{R}^{1_{+4}}\right) }
\end{array}
\\
& \downarrow \\
\underline{\mathcal{N}=2}: &
\mathcal{O}_{\mathcal{N}=2,STU,\mathbf{A3}\left[
\frac{1}{2}-BPS\right] },
\end{array}
\label{ddddd-1}
\end{equation}
where (recall Eq. (\ref{stu-1-charge}))
\begin{equation}
\mathcal{O}_{\mathcal{N}=2,STU,\mathbf{A3}\left[
\frac{1}{2}-BPS\right]
}\sim \mathcal{O}_{STU,doubly-crit}=\frac{\left[ \SL\left( 2,\mathds{R}%
\right) \right] ^{3}}{\left[ \SO\left( 1,1\right) \right] ^{2}\rtimes \mathds{%
R}^{3}}.  \label{ddddd-2}
\end{equation}
Consequently, Eqs. (\ref{ddddd-1})-(\ref{ddddd-2}) correspond to the
following chains of maximal symmetric embeddings: for the numerators it
holds (\ref{ch-1}), whereas for the stabilizers it holds ($\mathds{R}\times
\mathds{R}\times \mathds{R}\equiv \mathds{R}^{3}$):
\begin{eqnarray}
&&E_{6\left( 6\right) }\rtimes \mathds{R}^{27}  \notag \\
&\supsetneq &\SO\left( 5,5\right) \times \SO\left( 1,1\right) \rtimes \left(
\mathds{R}^{5,5_{-2}}\times \mathds{R}^{16_{+1}}\times \mathds{R}%
^{1_{+4}}\right)  \notag \\
&\supsetneq &\SO\left( 4,4\right) \times \SO\left( 1,1\right) \times \SO\left(
1,1\right)  \notag \\
&&\rtimes \left( \mathds{R}^{8_{v,-2,0}}\times \mathds{R}^{8_{c,+1,+1}}%
\times \mathds{R}^{8_{s,+1,-1}}\times \mathds{R}^{1_{-2,+2}}\times \mathds{R}%
^{1_{-2,-2}}\times \mathds{R}^{1_{+4,0}}\right)
\end{eqnarray}
where the subscript denote $\SO\left( 1,1\right) $-weights, and the
translational factors $\mathds{R}^{16_{+1}}$ and $\mathds{R}%
^{8_{v,-2,0}}\times \mathds{R}^{8_{c,+1,+1}}\times \mathds{R}^{8_{s,+1,-1}}$
are progressively truncated out.

Finally, it is worth remarking that a comparison of (\ref{1-charge-sol}) and
(\ref{nBPS}) explains the $\pi $-invariance characterizing both the $1$%
-charge solution (\ref{1-charge-sol}) and the attractor solution (\ref{nBPS}%
) with $\mathcal{I}_{4}<0$.

\subsection{\label{"extr"} \textit{``}Extremal'' $D=4$\ $STU$\ BHs}

It should be remarked that, consistent with the assumption made in
\cite {Bossard:2009at} and \cite{Bossard:2009we}, the extremality
characterizing the $D=4$ ``large'' and ``small'' BHs associated to
nilpotent $\SO_{0}\left( 4,4\right) $-orbits of real dimension $10$,
$12$, $16$ and $18$ (treated in Sects. \ref{Large-Extremal} and
\ref{Small-Extremal}) is an extremality \textit{which can be
obtained through a limit process} from a non-extremal BH solution.

However, there exist extremal $D=4$ BHs which cannot be seen as the
``extremal limit'' of non-extremal BH solutions. As done in \cite
{Borsten:2010db}, we dub them ``extremal'' BHs (i.e.\ with the quotation marks). An example of this
type of BHs is provided by the nilpotent $G_{3,t^{3}}\left(
=G_{2\left( 2\right) }\right) $-orbit $\mathcal{O}_{5}$. As given by
Table 1 and Fig. 2 (Hasse diagram of $G_{2\left( 2\right) }$, with
partial ordering relations) of \cite {Kim:2010bf}, this orbit is the
one with highest degree (namely, $7$; \textit{cfr.} Eq. (4.33) of
\cite{Kim:2010bf}) of nilpotency, and it is therein claimed not to
be given by the extremal limit of a non-extremal BH solution.
Through the (inverse of the) embedding procedure (\textit{cfr.} Eq.
(A.41) of \cite{Bossard:2009we})
\begin{equation}
\underset{G_{3,T^{3}}}{G_{2\left( 2\right) }}\subsetneq \underset{%
G_{3,ST^{2}}}{\SO_{0}\left( 4,3\right) }\subsetneq \underset{G_{3,STU}}{%
\SO_{0}\left( 4,4\right) },
\end{equation}
which is discussed at the end of App. A.3 of \cite{Bossard:2009we},
as well
as in Sect. 5 of \cite{Bellucci:2007zi}, the $G_{2\left( 2\right) }$-orbit $%
\mathcal{O}_{5}$ determines all nilpotent $\SO_{0}\left( 4,4\right)
$-orbits of real dimension larger than $18$, namely $20$, $22$ and
$24$, as resulting from Hasse diagram of $\SO_{0}\left( 4,4\right) $,
\textit{e.g.} given by Fig. 1 of \cite{Borsten:2010db}.

Since all nilpotent $G_{3}$-orbits are characterised by all the four $4$%
-qubit invariants vanishing (as resulting from page 14, Table 3 and
Table 6
of \cite{Chterental:2007}), the statement is that in general the $%
\SO_{0}\left( 4,4\right) $-nilpotent orbits of dimension $20$, $22$
and $24$ (respectively associated to the families $L_{a_{4}}$,
$L_{0_{5\oplus \overline{3}}}$ and $L_{0_{7\oplus \overline{1}}}$)
correspond to ``extremal'' (with $I_{1}=0$, ``small''
\textit{and/or} ``large'') $D=4$ BHs, which are \textit{not} the
limit of non-extremal BH solutions.

The $\SO_{0}\left( 4,4\right) $-nilpotent orbits corresponding to the
families $L_{ab_{3}}$, $L_{a_{4}}$, $L_{0_{5\oplus \overline{3}}}$ and $%
L_{0_{7\oplus \overline{1}}}$ generally contain $4$-way entangled
states. However, since we are considering $\SO_{0}\left( 4,4\right)
$-orbits which are nilpotent, the corresponding parameters
(\textit{if any}) are all set to zero (consistent with the claim at
page 14 of \cite{Chterental:2007}).

We leave the study of such \textit{``}extremal'' BHs for further
future investigation.

\section{Four-Way Entanglement of Eight
Qubits in $\mathcal{N}=8$ Supergravity...\label{sec:8qubit}}

Having now seen in some detail how the BHs of the $STU$ model are
intricately related to the entanglement of three and four qubits, it
is natural to ask whether this intriguing correspondence can be
extended to other supergravity theories. Given that the $STU$ model
may be embedded in the $\mathcal{N}=8$ theory this is a natural and
interesting case to consider, especially given its $E_{7(7)}$
U-duality group, which is rather exotic from the perspective of
quantum information theory. Indeed, the BHs of $\mathcal{N}=8$
supergravity were related to qubits in \cite
{Duff:2006ue,Levay:2006pt}. Since the BHs transform linearly under
the U-duality group $E_{7(7)}$ they cannot simply correspond to the
arbitrary entanglement of more qubits. Indeed, they are related to a
very special tripartite entanglement of seven qubits as described by
the Fano plane \cite {Duff:2006ue}.

The maximally supersymmetric $D=4,\mathcal{N}=8$ supergravity \cite
{Cremmer:1979up} is based on the degree-$3$
Jordan algebra $J_{3}^{\mathds{O}_{s}}$ of $3\times 3$ Hermitian matrices over the split form of the octonions $%
\mathds{O}_{s}$ \cite{Gunaydin:1983rk,Gunaydin:1983bi}. It  contains 70
scalar fields parametrising the coset (\ref {N=8-D=4-1}), where
$E_{7(7)}$ is the U-duality group and $\SU(8)$ its maximal compact
subgroup. There are also $28$ gauge potentials, which,
together with their $28$ magnetic duals, transform linearly as the $%
\ensuremath{\mathbf{56}}$ of $E_{7(7)}$. The stationary BH solutions
carry these charges and the extremal solutions have a
Bekestein-Hawking entropy given by
\begin{equation}
S=\pi \sqrt{\left| \mathcal{I}_{4}\right| },\label{entropyy}
\end{equation}
where $\mathcal{I}_{4}$ is the unique Cartan-Cremmer-Julia quartic
invariant of $E_{7(7)}$ \cite{Cartan,Cremmer:1979up} built from the 56
electromagnetic charges \cite{Kallosh:1996uy}.

\begin{figure}[ht]
  \centering
  \includegraphics[width=7cm]{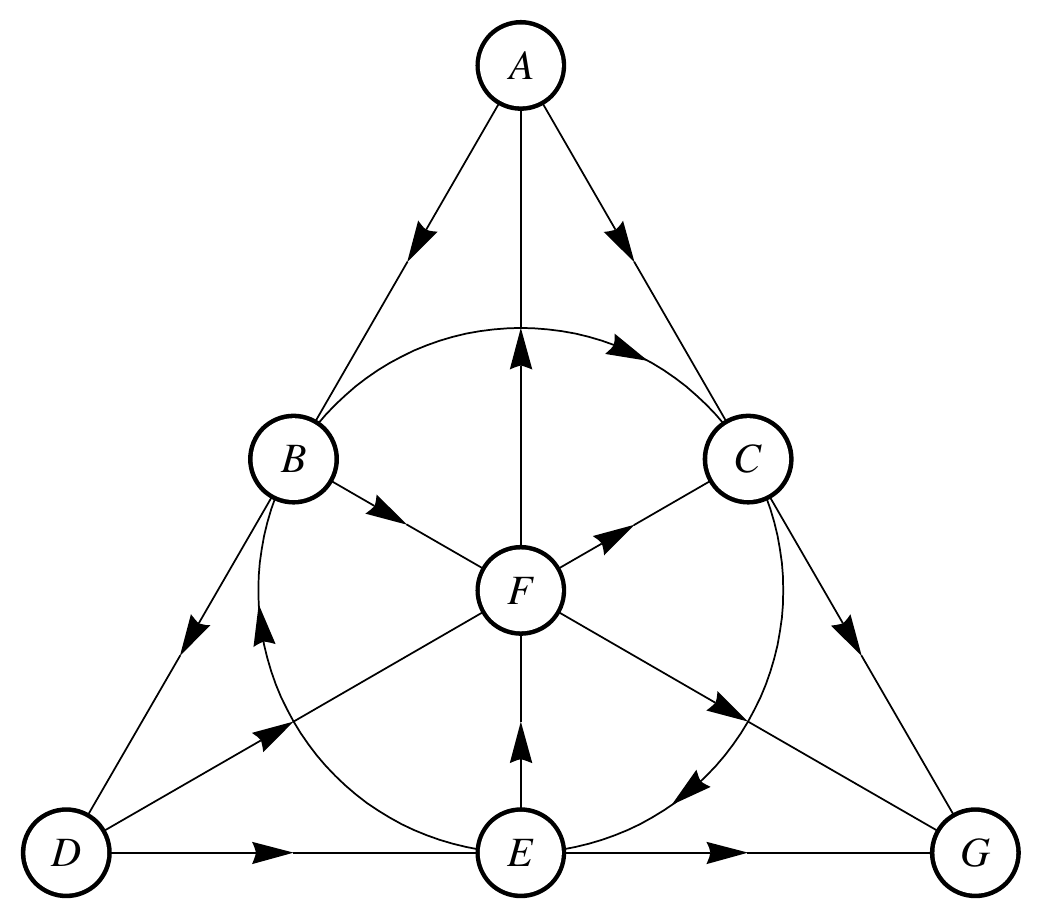}
  \caption[Fano plane]{The Fano plane is a projective plane with seven points and seven lines (the circle counts as a line). We may associate it to the state \eqref{eq:7QubitState} by interpreting the points as the seven qubits $A$-$G$ and the lines as the seven tripartite entanglements. }
  \label{fig:FanoPlane}
\end{figure}

The crucial observation relating the black holes to the tripartite entanglement of seven qubits is that $E_7$ contains seven copies of the single qubit SLOCC group $\SL(2)$ and that the $\mathbf{56}$ decomposes in a very particular way. Under
\be
\begin{array}{cccccccccccccccccccc}
E_{7(7)} &\supset& \SL(2)_A &\times &\SL(2)_B& \times& \SL(2)_C& \times &\SL(2)_D &\times &\SL(2)_E &\times &\SL(2)_F& \times &\SL(2)_G\\
\end{array}
\ee
the $\mathbf{56}$ decomposes as
\begin{equation}\label{eq:56Decomp}
\begin{split}
\mathbf{56} &\to\phantom{+\!}\mathbf{(2,2,1,2,1,1,1)} \\
&\phantom{\to}+\mathbf{(1,2,2,1,2,1,1)} \\
&\phantom{\to}+\mathbf{(1,1,2,2,1,2,1)} \\
&\phantom{\to}+\mathbf{(1,1,1,2,2,1,2)} \\
&\phantom{\to}+\mathbf{(2,1,1,1,2,2,1)} \\
&\phantom{\to}+\mathbf{(1,2,1,1,1,2,2)} \\
&\phantom{\to}+\mathbf{(2,1,2,1,1,1,2)}.
\end{split}
\end{equation}
Note, each term in the above decomposition transforms as a  $\mathbf{(2,2,2)}$ under three of the $\SL(2)$ factors and as singlets under the remaining four, but taken together they transform as the $\mathbf{56}$ of $E_{7(7)}$.  This translates into seven intertwined copies of the 3-qubit Hilbert space:
\begin{equation}\label{eq:7QubitState}
\begin{array}{c@{}c@{\ }c@{\lvert}*{3}{@{}c}@{\rangle}c}
\ket{\Psi}_{56} = &   & a_{ABD} & A       & B       & D \\
                  & + & b_{BCE} & B       & C       & E \\
                  & + & c_{CDF} & C       & D       & F \\
                  & + & d_{DEG} & D       & E       & G \\
                  & + & e_{EFA} & E       & F       & A \\
                  & + & f_{FGB} & F       & G       & B \\
                  & + & g_{GAC} & G       & A       & C &.
\end{array}
\end{equation}
This state has a very distinctive structure:
\begin{enumerate}
\item Two distinct qubits appear together in one and only one tripartite
entanglement.
\item Any two tripartite entanglements have at least one qubit in common.
\item Every qubit belongs to three distinct tripartite entanglements.
\end{enumerate}
On replacing the words qubit and tripartite entanglement with the words
point and line, respectively, it becomes apparent that the state describes
the projective plane of order 2. This is know as the Fano plane, which is depicted
in \autoref{fig:FanoPlane}.
The Fano plane is also the multiplication table of the imaginary
octonions. This special state has revealed a number of interesting
connections relating exceptional groups, octonions and special
finite geometries to quantum information  theory and, in particular,
three qubits
\cite{Levay:2008mi,Borsten:2008wd,Cerchiai:2010tk,Levay:2009bp}.
\medskip

Repeating the analysis of the $STU$ model for the maximally
supersymmetric theory leads to another exotic qubit configuration:
the four-way entanglement of eight qubits. A time-like reduction of
the $\mathcal{N}=8$
theory yields the scalar manifold given in (\ref{N=8-D=3-1}), where $%
E_{8(8)}$ is the $D=3$ U-duality group and $\SO^{\star }(16)$ is a
non-compact form of its maximal compact subgroup. The stationary BH
solutions are given by the geodesics in (\ref{N=8-D=3-1}) that are
in turn parametrised by $\ensuremath{\mathfrak{e}}_{8(8)}$ valued
Noether charges.
In particular, the extremal solutions correspond to the nilpotent orbits of $%
E_{8(8)}$ acting on $\ensuremath{\mathfrak{e}}_{8(8)}$. The Cartan
decomposition \be \alg{e}_{8(8)}=\alg{so}^*(16)\oplus \rep{128}, \ee
where $ \rep{128}$ is the spinor of $\SO^\star(16)$, implies, by the
Kostant-Sekiguchi theorem, that the orbits the nilpotent  orbits of
$E_{8(8)}$ acting on $\alg{e}_{8(8)}$ are in one-to-one
correspondence to the nilpotent orbits of  $\SO(16, \field{C})$
acting on $\rep{128}$
\cite{Bergshoeff:2008be,Bossard:2009we,Bossard:2009at}. The 128
independent components are given by the  $28+28$ electromagnetic
charges, the NUT charge, the mass and 70 scalars of the
$\mathcal{N}=8$ theory.

The qubit interpretation is obtained by decomposing the adjoint (fundamental) of $E_{8(8)}$ with respect to $[\SL(2)]^8$ \cite{Borsten:2008wd}. This can be interpreted as the  time-like reduction of the tripartite entanglement of seven qubits, the eighth $\SL(2)$ being the Ehlers group. Explicitly,
\begin{gather}\label{fri}
E_8 \supset \SL(2)_A \times \SL(2)_B \times \SL(2)_C \times \SL(2)_D
\times \SL(2)_E \times \SL(2)_F \times \SL(2)_G  \times \SL(2)_H,
\shortintertext{under which}
\begin{split}\label{248}
\mathbf{248} &\to\phantom{+\!}\mathbf{(3,1,1,1,1,1,1,1)+(2,2,2,1,2,1,1,1)+(1,1,1,2,1,2,2,2)} \\
&\phantom{\to}+\mathbf{(1,3,1,1,1,1,1,1)+(2,1,2,2,1,2,1,1)+(1,2,1,1,2,1,2,2)} \\
&\phantom{\to}+\mathbf{(1,1,3,1,1,1,1,1)+(2,1,1,2,2,1,2,1)+(1,2,2,1,1,2,1,2)} \\
&\phantom{\to}+\mathbf{(1,1,1,3,1,1,1,1)+(2,1,1,1,2,2,1,2)+(1,2,2,2,1,1,2,1)} \\
&\phantom{\to}+\mathbf{(1,1,1,1,3,1,1,1)+(2,2,1,1,1,2,2,1)+(1,1,2,2,2,1,1,2)} \\
&\phantom{\to}+\mathbf{(1,1,1,1,1,3,1,1)+(2,1,2,1,1,1,2,2)+(1,2,1,2,2,2,1,1)} \\
&\phantom{\to}+\mathbf{(1,1,1,1,1,1,3,1)+(2,2,1,2,1,1,1,2)+(1,1,2,1,2,2,2,1)} \\
&\phantom{\to}+\mathbf{(1,1,1,1,1,1,1,3)},
\end{split}
\end{gather}
and
\be
\alg{e}_{8(8)}\cong [\alg{sl}(2, \R)]^8 \oplus \alg{p}
\ee
where
\be
\begin{split}\label{p-Frak}
\alg{p}=   &~\phantom{+}~\mathbf{(2,2,2,1,2,1,1,1)+(1,1,1,2,1,2,2,2)} \\
&+\mathbf{(2,1,2,2,1,2,1,1)+(1,2,1,1,2,1,2,2)} \\
 &+\mathbf{(2,1,1,2,2,1,2,1)+(1,2,2,1,1,2,1,2)} \\
 &+\mathbf{(2,1,1,1,2,2,1,2)+(1,2,2,2,1,1,2,1)} \\
&+\mathbf{(2,2,1,1,1,2,2,1)+(1,1,2,2,2,1,1,2)} \\
 &+\mathbf{(2,1,2,1,1,1,2,2)+(1,2,1,2,2,2,1,1)} \\
&+\mathbf{(2,2,1,2,1,1,1,2)+(1,1,2,1,2,2,2,1)},
\end{split}
\ee
which admits an interpretation as the four-way entanglement of eight qubits,
\be
\begin{array}{c@{}c@{\ }c@{\lvert}*{8}{@{}c}@{\rangle\ +\ }c@{\lvert}*{8}{@{}c}@{\rangle}@{}c}\label{eq:ourstste?}
\ket{\Psi}_{224} = &   & a_{HABD}         & H       & A       & B       & \bullet & D       & \bullet & \bullet & \bullet
                       & \tilde{a}_{CEFG} & \bullet & \bullet & \bullet & C       & \bullet & E       &       F & G       \\
                   & + & b_{HBCE}         & H       & \bullet & B       & C       & \bullet & E       & \bullet & \bullet
                       & \tilde{b}_{DFGA} & \bullet & A       & \bullet & \bullet & D       & \bullet &       F & G       \\
                   & + & c_{HCDF}         & H       & \bullet & \bullet & C       & D       & \bullet &       F & \bullet
                       & \tilde{c}_{EGAB} & \bullet & A       & B       & \bullet & \bullet & E       & \bullet & G       \\
                   & + & d_{HDEG}         & H       & \bullet & \bullet & \bullet & D       & E       & \bullet & G
                       & \tilde{d}_{FABC} & \bullet & A       & B       & C       & \bullet & \bullet &       F & \bullet \\
                   & + & e_{HEFA}         & H       & A       & \bullet & \bullet & \bullet & E       &       F & \bullet
                       & \tilde{e}_{GBCD} & \bullet & \bullet & B       & C       & D       & \bullet & \bullet & \bullet \\
                   & + & f_{HFGB}         & H       & \bullet & B       & \bullet & \bullet & \bullet &       F & G
                       & \tilde{f}_{ACDE} & \bullet & A       & \bullet & C       & D       & E       & \bullet & \bullet \\
                   & + & g_{HGAC}         & H       & A       & \bullet & C       & \bullet & \bullet & \bullet & G
                       & \tilde{g}_{BDEF} & \bullet & \bullet & B       & \bullet & D       & E       &       F & \bullet & .
\end{array}
\ee
Half the states are given by the  quadrangles of the Fano plane and
the other half by the quadrangles of the dual Fano plane. See also
\cite{Manivel:2005}, where this configuration was related to a \emph{doubled Fano plane}.
While we may assign $|\Psi \rangle _{224}$ to the coset\footnote{%
Note that $E_{8\left( 8\right) }$ contains (by a chain of maximal
symmetric embeddings) $\left[ \SL\left( 2,\mathds{R}\right) \right]
^{8}$, $\left[
\SL\left( 2,\mathds{R}\right) \right] ^{4}\times \left[ \SU\left( 2\right) %
\right] ^{4}$, and $\left[ \SU\left( 2\right) \right] ^{8}$. These
correspond
to different completions of the $8$-dimensional Cartan subalgebra of $%
E_{8\left( 8\right) }$. However, if one constrains the first
inclusion of the chain to be $E_{8\left( 8\right) }\supset
E_{7\left( 7\right) }\times \SL(2,\mathds{R})$, then the embedding of
$\left[ \SU\left( 2\right) \right]
^{8}$ is excluded.} $E_{8(8)}/[\SL(2,%
\mathds{R})]^{8}$, unlike the $STU$ example, the Kostant-Sekiguchi
theorem does \textit{not} apply. Indeed,
$E_{8(8)}/[\SL(2,\mathds{R})]^{8}$ is not a symmetric space, as it
can be verified \textit{e.g.} by considering the non-zero
commutation relations of, for example, two elements in
$\mathbf{(2,2,2,1,2,}$ $\mathbf{1,1,1)}$ and
$\mathbf{(2,1,2,2,1,2,1,1)}$. This is in fact as one would
anticipate, since in four dimensions the tripartite entanglement of
seven qubits forms a representation of the full U-duality group,
not just its $[\SL(2)]^{7}$ subgroup. Consequently, performing the
time-like reduction, it is actually the nilpotent orbits of
$\SO(16,\mathds{C})$ acting on its spinorial representation that are
of relevance.

\section{\label{sec:8qubit-exc}...and in $\mathcal{N}=2$ Exceptional Supergravity}

Another interesting case to consider is the $\mathcal{N}=2$
exceptional supergravity, namely the magic model based on the
degree-$3$ Euclidean
Jordan algebra $J_{3}^{\mathds{O}}$ of $3\times 3$ Hermitian matrices over the division algebra of  octonions $%
\mathds{O}$ \cite{Gunaydin:1983rk,Gunaydin:1983bi}. This is the
unique magic model which cannot be obtained as consistent truncation
of the maximal theory treated in previous Section. The extension of
the connection between BHs and QIT to the case of magic supergravity
was firstly suggested by Levay in \cite{Levay:2006pt} (based on work
of \cite{Manivel:2005}\ and \cite {Elduque:2005}), and it has been
considered by one of the present authors and Ferrara in
\cite{Duff:2007wa}. The U-duality group of the $D=4$
exceptional theory is another non-compact, real form of $E_{7}$, namely $%
E_{7\left( -25\right) }$, which, as its maximal counterpart
$E_{7\left( 7\right) }$, is rather exotic from a QIT perspective, as
well. Once again, since the BHs transform linearly under the
U-duality group $E_{7(-25)}$, they are not expected to be related to
an arbitrary entanglement of more qubits. As it will become evident
from treatment below, after a timelike reduction to $D=3$ they
result to be related to a curious combination of local unitary and
special linear factor groups within an entanglement of eight qubits
with the same structure of the one treated in previous Section.

The exceptional magic $D=4,$ $\mathcal{N}=2$ supergravity \cite
{Gunaydin:1983rk,Gunaydin:1983bi} has 27 complex scalar fields (one
for each Abelian vector multiplet), parametrising the rank-$3$
special K\"{a}hler symmetric coset
\begin{equation}
\frac{E_{7\left( -25\right) }}{E_{6\left( -78\right) }\times \Un\left(
1\right) },  \label{N=2-exc-D=4-1}
\end{equation}
where $E_{7\left( -25\right) }$ is the U-duality group and
$E_{6\left(
-78\right) }\times \Un\left( 1\right) $ its maximal compact subgroup. There $%
28=1$ (graviphoton)$+27$ gauge potentials, together with their $28$
magnetic duals, transform linearly as the $\ensuremath{\mathbf{56}}$
of $E_{7\left( -25\right) }$. The stationary BH solutions carry
these charges and the extremal solutions have a Bekestein-Hawking
entropy given by the same formula (\ref{entropyy}) of the
$\mathcal{N}=8$ case, where $\mathcal{I}_{4}$ is now the unique
quartic invariant of $E_{7(-25)}$ built of the 56 electromagnetic
charges.

When considering the groups in the complex field, \textit{mutatis
mutandis} the story goes as in the maximal theory treated in
previous Section, but the intepretation in terms of timelike
reduction down
to $D=3$ is different. Indeed, by performing such a reduction, the $\mathcal{%
N}=2$ exceptional theory yields the scalar manifold to become the
rank-$4$ para-quaternionic, pseudo-Riemannian symmetric coset
\begin{equation}
\frac{E_{8\left( -24\right) }}{E_{7\left( -25\right) }\times \SL\left( 2,%
\mathds{R}\right) },  \label{N=2-exc-D=3-1}
\end{equation}
which is obtained from (\ref{N=2-exc-D=4-1}) through the so-called
$c^{\ast } $\textit{-map} (\cite{Cecotti:1988qn},
\cite{Bergshoeff:2008be}, and Refs. therein). $E_{8(-24)}$ is the
$D=3$ U-duality group and $E_{7\left( -25\right) }\times \SL\left(
2,\mathds{R}\right) $ is a non-compact form of its maximal compact
subgroup, the factor $\SL\left( 2,\mathds{R}\right) $ being the
Ehlers group\footnote{Note that $E_{8\left( -24\right)}$ contains
(in a maximal and symmetric way, see \textit{e.g.}
\cite{Helgason:1978}) $\SO^\star(16)$, but the $c^{\ast }
$\textit{-map} determines the
relevant subgroup to be $E_{7\left( -25\right) }\times \SL\left( 2,%
\mathds{R}\right)$. From a physical perspective, this can ultimately
be related to the split between the gravity and vector multiplets,
which is not present in the maximal theory treated in previous
Section.} The stationary BH solutions are given by the
geodesics in (\ref{N=2-exc-D=3-1}) that are in turn parametrised by $%
\ensuremath{\mathfrak{e}}_{8(-24)}$ valued Noether charges. In
particular, the extremal solutions correspond to the nilpotent
orbits of $E_{8(-24)}$ acting on
$\ensuremath{\mathfrak{e}}_{8(-24)}$. The Cartan decomposition
\begin{equation}\label{CERN-1}
\frak{e}_{8(-24)}=\left( \frak{e}_{7\left( -25\right) }+\frak{sl}\left( 2,%
\mathds{R}\right) \right) \oplus \left(
\mathbf{56},\mathbf{2}\right) ,
\end{equation}
implies, by the Kostant-Sekiguchi theorem, that the nilpotent orbits of $%
E_{8(-24)}$ acting on $\ensuremath{\mathfrak{e}}_{8(-24)}$ are in
one-to-one correspondence to the nilpotent orbits of $E_{7}\left(
\mathds{C}\right)
\times \SL\left( 2,\mathds{C}\right) $ acting on $\left( \mathbf{56},\mathbf{2%
}\right) $. In the real field, these latter has $112$ independent
components, given by the $28+28$ electromagnetic charges, the NUT
charge, the mass and 54 (27 complex) scalar degrees of freedom of
the $\mathcal{N}=2$, $D=4$ exceptional theory.

The qubit interpretation is obtained by decomposing the adjoint
(fundamental) of $E_{8(-24)}$ with respect to $[\SL(2,\mathds{R})]^{4}\times %
\left[ \SU\left( 2\right) \right] ^{4}$. Indeed, differently from
$E_{8\left( 8\right) }$, $E_{8(-24)}$ contains (by a chain of
maximal symmetric embeddings) $\left[ \SL\left( 2,\mathds{R}\right)
\right] ^{4}\times \left[ \SU\left( 2\right) \right] ^{4}$ and
$\left[ \SU\left( 2\right) \right] ^{8}$ (corresponding to different
completions of the $8$-dimensional Cartan
subalgebra of $E_{8\left( -24\right) }$), but not $\left[ \SL\left( 2,\mathds{%
R}\right) \right] ^{8}$.  If one further constrains the first
inclusion of the chain to be the relevant one for $c^{\ast }$-map
$E_{8\left( -24\right) }\supset E_{7\left( -25\right) }\times
\SL(2,\mathds{R})$, then also the embedding of $\left[\SU\left(
2\right) \right] ^{8}$ is excluded. Explicitly, the decompositions
(in the complex field) (\ref{fri}) and (\ref {248}) still hold, but
the time-like reduction interpretation is different, because it here
concerns the tripartite entanglement of seven qubits which are
split, on the real field, into four qubits transforming under
$\SU\left( 2\right) $ and three qubits transforming under $\SL(2,\mathds{R})$ qubits, the fourth $%
\SL(2,\mathds{R})$ being the Ehlers group (the very same commuting
with $E_{7\left( -25\right)}$ inside $E_{8\left( -24\right)}$, see
(\ref{CERN-1})). In this case, it holds that
\begin{equation}
\frak{e}_{8(-24)}\cong \lbrack \left( \frak{sl}\left(
2,\mathds{R}\right) ]^{4}+[\frak{su}\left( 2\right) ]^{4}\right)
\oplus \frak{p},
\end{equation}
where $\frak{p}$ has the same formal decomposition as given in (%
\ref{p-Frak}), but with the second quartet of irreprs. pertaining to
$\left[
\SU\left( 2\right) \right] ^{4}$, and not to $\left[ \SL\left( 2,\mathds{R}%
\right) \right] ^{4}$ as in the maximal case. This admits an
interpretation as the four-way entanglement of eight qubits,
democratically covariant with respect to the two possible symmetry
groups of SLOCC-equivalent real qubits, namely four with respect to $\SU\left( 2\right) $ and four with respect to $\SL(2,\mathds{R%
})$. This is a rather weird split combination from the QIT point of
view, and we leave for future investigation the question of whether
this setup enjoys any real use.

While one can formally still assign $|\Psi \rangle _{224}$ to the
coset
\begin{equation}
\frac{E_{8(-24)}}{[\SL(2,\mathds{R})]^{4}\times \left[ \SU\left( 2\right) %
\right] ^{4}},\label{coss}
\end{equation}
unlike the $STU$ example and analogously to the $\mathcal{N}=8$ case
treated in previous Section, the Kostant-Sekiguchi theorem does
\textit{not} apply,
because (\ref{coss}) is not a symmetric space (as it can be verified \textit{%
e.g.} by considering the non-zero commutation relations of, for
example, two elements in $\mathbf{(2,2,2,1,2,1,1,1)}$ and
$\mathbf{(2,1,2,2,1,2,1,1)}$). Once again, this can be traced back
to the mismatching between the $[\SL(2)]^{7}$ group and the whole
$D=4$ U-duality group (in the complex field); indeed, in $D=4$ the
tripartite entanglement of seven qubits forms a representation of
the full U-duality group, not just its $[\SL(2)]^{7}$ subgroup.
Consequently, performing the time-like reduction, it is actually the
nilpotent orbits of $E_{7}\left( \mathds{C}\right) \times \SL\left( 2,\mathds{%
C}\right) $ acting on the irrepr. $\left(
\mathbf{56},\mathbf{2}\right) $ that are of relevance.

\section*{Acknowledgements}
We would like to thank Sergio Ferrara and Philip Gibbs for useful conversations. The work of LB is supported by ERC advanced grant no. 226455, \emph{Supersymmetry, Quantum Gravity and Gauge Fields} (SUPERFIELDS).  The work of MJD is supported by the STFC under rolling grant ST/G000743/1.


\providecommand{\href}[2]{#2}\begingroup\raggedright\endgroup

\end{document}